\documentclass[preprint,1p,times,authoryear]{elsarticle}

\usepackage{amssymb}
\usepackage{amsthm}

\usepackage{natbib}
\usepackage[utf8]{inputenc} 
\usepackage[T1]{fontenc}    
\usepackage{url}            
\usepackage{booktabs}       
\usepackage{amsfonts}       
\usepackage{nicefrac}       
\usepackage{microtype} 
\usepackage{hyperref}
\hypersetup{
  colorlinks,
  citecolor=violet,
  linkcolor=red,
  urlcolor=blue,
}
\usepackage{subcaption}
\usepackage{enumitem}

\usepackage[misc,geometry]{ifsym}

\usepackage{verbatim}
\usepackage{array}
\usepackage{color}
\usepackage{caption}
\usepackage{multirow}
\usepackage[table]{xcolor}

\usepackage{csquotes}

\usepackage{pdflscape}

\usepackage{censor}

\journal{}

\begin{document}


\renewcommand{\figureautorefname}{Figure }
\newcommand{\figuresautorefname}{Figures }
\renewcommand{\tableautorefname}{Table }
\renewcommand{\equationautorefname}{Equation }
\newcommand{\equationsautorefname}{Equations }

\renewcommand{\sectionautorefname}{Section}
\renewcommand{\subsectionautorefname}{Section }
\renewcommand{\subsubsectionautorefname}{Section }		
\newcommand{\sectionsautorefname}{Sections }

\begin{frontmatter}



\title{On Implementing Autonomous Supply Chains: a Multi-Agent System Approach\tnoteref{tlabel}}
\tnotetext[tlabel]{This work is an extension of an IFAC 2023 conference paper. This word has been accepted to Computers in Industry for publication.}

\author{Liming Xu\fnref{inst1}}
\ead{lx249@cam.ac.uk}
\author[inst1]{Stephen Mak}
\ead{sm2410@cam.ac.uk}
\author{Maria Minaricova\fnref{inst2}}
\ead{maria.minaricova@fetch.ai}
\author{Alexandra Brintrup\fnref{inst1}}
\ead{ab702@cam.ac.uk}

\affiliation[inst1]{organization={Institute for Manufacturing,
                                  Department of Engineering, 
                                  University of Cambridge},
                    addressline={17 Charles Babbage Road}, 
                    city={Cambridge},
                    postcode={CB3 0FS}, 
                    country={United Kingdom}}

\affiliation[inst1]{organization={Fetch.ai},
                    addressline={St John's Innovation Centre, Cowley Rd}, 
                    city={Cambridge},
                    postcode={CB4 0WS}, 
                    country={United Kingdom}}

\begin{abstract}
Trade restrictions, the COVID-19 pandemic, and geopolitical conflicts have significantly exposed vulnerabilities within traditional global supply chains. 
These events underscore the need for organisations to establish more resilient and flexible supply chains. 
To address these challenges, the concept of the autonomous supply chain (ASC), characterised by predictive and self-decision-making capabilities, has recently emerged as a promising solution.
However, research on ASCs is relatively limited, with no existing studies specifically focusing on their implementations.
This paper aims to address this gap by presenting an implementation of ASC using a multi-agent approach. 
It presents a methodology for the analysis and design of such an agent-based ASC system (A2SC).
This paper provides a concrete case study, the autonomous meat supply chain, which showcases the practical implementation of the A2SC system using the proposed methodology.
Additionally, a system architecture and a toolkit for developing such A2SC systems are presented. 
Despite limitations, this work demonstrates a promising approach for implementing an effective ASC system.
\end{abstract}

\begin{keyword}
Autonomous supply chain 
\sep Multi-agent system
\sep Autonomous agents
\sep Perishable foods
\end{keyword}

\end{frontmatter}


\section{Introduction}\label{sec:introduction}
Trade restrictions, the COVID-19 pandemic, and geopolitical conflicts have seriously exposed vulnerabilities within traditional global supply chains \citep{handfield2020corona, shih2020global}, highlighting the need for organisations to establish more resilient and flexible supply chains. 
Indeed, traditional supply chains exhibit limited agility and resilience, with over 70\% of surveyed supply chain leaders acknowledging the need for enhanced agility and operational resilience \citep{nelsonhall2021moving}. 
The model of autonomous supply chains (ASC) has been proposed as a concept solution to achieve supply chain resilience and structural flexibility \citep{xu2023autonomous}.

Traditional supply chains still heavily rely on manual processes, which can often introduce communication delays and human errors, ultimately leading to increased costs \citep{nelsonhall2021moving}. 
Intelligent digital technologies, such as AI and robotics, are penetrating into the supply chain domain, automating various processes, particularly those error-prone and routine tasks.
However, these automation efforts mainly focus on individual processes, with limited integration between processes, let alone inter-firm integration. 
The needs for better supply chain performance require more horizontal and vertical integration \citep{frohlich2001arcs,stevens2016integrating}. 
ASCs emphasise interconnection, integration, automation, and self-decision-making. 
Within this framework, the supply chain has capacity to predict disruptions and automatically adapt to changes through reconfiguration and adjustments. 
Consequently, ASCs offer a promising solution to address challenges in an uncertain and turbulent business environment.

The concept of supply chains equipped with such predictive and self-adaptive capabilities have been a topic of discussion for many years. 
A ``smarter'' supply chain, capable of autonomous learning and decision-making without human intervention, was proposed by \citet{butner2010smarter}. 
This envisioned supply chain was characterised as instrumented, interconnected, and intelligent.  
\citet{wu2016smart} conducted a literature review on the impact of digital technologies on supply chain management (SCM) and proposed three additional defining characteristics for this emerging supply chain: automation, integration, and innovation.
Expanding on these ideas, \citet{calatayud2019self} proposed the concept of the {\it self-thinking} supply chain, which is distinguished by autonomous and predictive capabilities relying on IoT and AI-enabled connectivity.
Moreover, \citet{xu2023autonomous} extended these concepts by formally proposing the autonomous supply chain model.
This work extensively delved into the theoretical aspects of ASCs, including their formal definition, defining characteristics, and a set of auxiliary concepts. 
Instead of solely outlining the characteristics of this intelligent supply chain, it presented a five-layer conceptual model and a seven-level supply chain autonomy reference model. 
These conceptual models can be used for providing guidelines for implementing ASC systems.

However, despite the existence of initial concepts and theoretical studies in the literature, the development of ASCs remains in its early stages.
A supply chain consists of a network of business entities engaged in various activities that transform raw materials into finished products and distribute them to customers \citep{christopher2016logistics}. 
These activities include demand forecast, supplier selection, material procurement, inventory management, and order fulfilment. 
Effectively managing the flow of goods and its associated information and financial flows is crucial for SCM.
To achieve autonomous SCM, interoperability between independent and geographically distributed entities must be facilitated. 
The multi-agent system (MAS) approach, which achieves distributed problem-solving through multiple intelligent agents interacting within a shared environment, are inherently well-suited for architecturally modelling supply chains \citep{wooldridge1995intelligent,fox2001agent, xu2023implementation}. 
While the MAS approach has been long recognised by academics and practitioners for its potential to automate and autonomise decision-making in SCM, its applications remain limited mainly to addressing the automation of specific functions or processes \citep{xu2021will, nitsche2023impact}.

For instance, \citet{ameri2013multi} introduced an agent-based digital manufacturing marketplace, enabling service consumers to autonomously search for and connect with corresponding service providers based on a predefined similarity metric. 
Despite facilitating the creation of configurable supply chains, this service matchmaking still targets the automation of a single function --- supplier selection.
In a more recent study, \citet{nitsche2023impact} conducted interviews with a cohort of  supply chain experts, identifying eleven MAS implementation use cases in SCM, all primarily centred around automating specific functions rather than the automation of physical flows and their associated information flows.
Furthermore, these use cases heavily rely on agent simulation/development platforms such as JADE \citep{bellifemine2005jade} for testing and evaluation, which deviates from realistic settings and exhibits limited compatibility with legacy systems \citep{xu2021will}.
Consequently, this reliance contributes to the slow industrial adoption of MAS for automation and autonomisation in SCM.

Additionally, current research rarely focuses on automating the entire SCM, which involves managing the three essential flows from material procurement to the final delivery of finished product to customers.
Autonomous management of these three flows is crucial for achieving ASCs \citep{xu2023autonomous}.
This paper, therefore, aims to bridge this gap by systematically exploring the development of ASCs with integrated processes.
While \citet{xu2023autonomous} presented a brief overview of an ASC implementation case, this paper delves comprehensively into all facets of implementing an ASC system, employing a multi-agent system (MAS) approach. 
Compared to existing studies, this work distinguishes itself by 
1) adopting a software engineering methodology --- Gaia \citep{wooldridge2000gaia} and its variant ROADMAP \citep{juan2002roadmap} --- as system design and analysis approaches; 
2) tackling the automation of the entire process from sourcing to delivery to customers; 
and 3) implementing a software system prototype rather than a simulation. 
Specifically, this work investigates the development of an agent-based ASC system, called A2SC, that facilitates interoperability and automation among distributed supply chain entities.

In this paper, we scrutinise the fundamental issues relevant to the design and development of a {\it practical} A2SC system, including its analysis, design, and implementation.
The main contributions of this paper can be summarised as follows:
\begin{itemize}[nosep]
  \item We investigate the application of the MAS approach in automating integrated processes and introduce an agent-based ASC.
  \item We present a methodology designed for A2SC analysis and design, and detail the realisation of A2SC systems using this methodology.
  \item We detail a concrete case study --- the autonomous meat supply chain --- demonstrating the technical implementation of an A2SC system.
  \item We report a system architecture and a toolkit for A2SC prototype development, providing guidance for implementing similar decentralised systems. 
\end{itemize}

The rest of this paper is organised as follows. 
\autoref{sec:related_work} reviews related work. 
\autoref{sec:a2sc} presents A2SCs and a methodology for their analysis and design. 
\autoref{sec:a2sc_implementation} details an A2SC implementation case study.
\autoref{sec:discussion} discusses the limitation and implications of this work. 
Finally, \autoref{sec:conclusion} concludes this paper and discusses future work.

\section{Related Work}\label{sec:related_work}
While ASCs are still in their early stage, there exists a body of research pertinent to their development.
We classify these work into two categories: conceptual discussions focusing on the transition towards ASCs and related technical attempts aiming at automating and autonomising SCM through MAS.
In this section, we provide a review of these two categories of related work.

\subsection{Transition to ASCs}
The initial discussion of ASCs can be traced back the concept of the ``smarter supply chain'' proposed by \citet{butner2010smarter} --- a supply chain capable of autonomous learning and decision-making without human intervention.
Despite the description of the characteristics of such supply chains in this article, their conceptual or theoretical aspects of ASCs have not been extensively studied.
\citet{wu2016smart} further developed the concept and investigated various technologies relevant to achieving a ``smart supply chain''. 
Built upon both physical and information connectivity \citep{calatayud2017connected}, \citet{calatayud2019self} proposed the ``self-think supply chain'', characterised by autonomous and predictive capabilities. 
Similar to \citet{wu2016smart}, this work defines the characteristics of this type of supply chains and identifies key enabling technologies to achieve them.
\citet{nitsche2021exploring, nitsche2021application} proposed a conceptual framework that specifies application areas and antecedents for realising automation in SCM and logistics.
Compared to previous studies, this work concretises concepts and centres around conceptualisation of the automation of physical and information processes.
Although this conceptual framework provides a common basis for further discussion on automation between research and practice, it falls short of guiding the technical development of ASCs.

\citet{xu2023autonomous} delved deeper into the conceptualisation of ASCs, proposing a set of conceptual artefacts, including structural entities and the MIISI model, to guide ASC implementation.
As described in \citet{nitsche2021exploring, xu2023autonomous}, fully automating a supply chain involves replacing manual processes by automated ones and establishing connections between these automated processes.
These automated processes served as the building blocks for ASCs and can be managed through structural entities. 
These structural entities are entities along a supply chain network that assemble and control the flows of materials, information, and finance.
They are responsible for managing the three flows that traverse these interconnected processes.
In A2SCs, these structural entities can be represented by software agents that autonomously manage their internal units and interacting with external entities.
All these works focus on conceptual aspects of ASCs; however, none of them directly deal with their technical implementation. 
We review the work related to their technical implementation in the next section.

\subsection{Realisation of ASCs}
A supply chain process requires numerous decisions among a multitude of actors across a supply chain.
The MAS approach, which involves a collection of software agents interacting with each other to solve a complex problem, is therefore well-suited to facilitate the automation and autonomisation of these processes \citep{dorri2018multi, xu2021will, nitsche2023impact}.
Since its initial adoption in SCM \citep{fox1993integrated}, yet still under-researched \citep{dorri2018multi}, this approach has been widely explored in this domain.

These applications can be broadly grouped into four categories: agent frameworks, dynamic supply chain formation and configuration, coordination and negotiation, and enhancement of specific supply chain functions.
The first category focuses on identifying common characteristics of generic or specific supply chains, providing an architectural framework for constructing agent-based SCM. 
For instance, \citet{sadeh2001mascot} introduced a MASCOT architecture based on the blackboard architecture \citep{erman1980hearsay}. 
\citet{julka2002agent} defined a unified framework for modelling the entire supply chain network in an object-oriented manner, representing supply chain entities as agents and materials and information represented as objects. 
\citet{lou2004study} proposed an agent architecture consisting of mediator and functional agents for agile SCM.
Additionally, agent frameworks designed for supply chains of specific industries, such as fashion \citep{lo2008framework}, wine \citep{kumar2013multi}, and outsourcing SME manufacturing \citep{kumar2013multi}, have also been proposed.

The second category utilises the MAS approach to address dynamic supply chain formation and configuration. 
\citet{chen1999negotiation} introduced an MAS that employs a set of performatives for pair-wise negotiation, facilitating dynamic supply chain generation in an open environment through automated or semi-automated negotiation. 
\citet{akanle2008agent} developed an agent-based model to manage customer demand by optimising supply chain configuration.
\citet{kim2010supply} investigated the dynamic supply chain formation problem based on the single machine earliness/tardiness model and utilised agent negotiation to address this issue in a three-tier supply chain setting. 
Meanwhile, \citet{ameri2013multi} presented a digital manufacturing market for matching service providers and consumers, enabling semantic matchmaking and thereby facilitating supply chain reconfiguration.

The third category specifically addresses coordination and negotiation within SCM.
While the second category focuses on the structural aspects of supply chains, this category deals with their dynamical aspects \citep{swaminathan2003models} --- coordination and negotiation.
\citet{barbuceanu1996coordinating} addressed the coordination problem by explicitly representing interaction knowledge among agents and introduced COOL, a structured conversion-based communication language for supply chain coordination. 
\citet{sadeh2001mascot} developed the MASCOT architecture, which provides protocols for both horizontal and vertical coordination in supply chains. 
\citet{xue2005agent} proposed an agent framework for coordinating construction supply chains, leveraging multi-issue negotiation using multi-attribute utility theory (MAUT). 
\citet{wong2010multi} presented a contract-net-like multilateral and multi-issue protocol designed for handling buyers-seller bargaining and interaction negotiation in SCM. 
This protocol also employs the MAUT approach to model utility functions during negotiations. 
Additionally, various general negotiation approaches suitable for SCM have been proposed, including the approach by \citet{chen2015approach}.

The last category, which is main focus of most literature, aims to improve specific supply chain functions or processes through the MAS approach. 
These include enhancing demand forecast \citep{liang2006agent}, 
managing costs \citep{fu2015adaptive}, 
selecting suppliers \citep{ghadimi2018multi, ameri2013multi, yu2015multi}, 
scheduling \citep{aminzadegan2019multi}, 
planning and controlling inventory \citep{ito2002agent, kazemi2009multi, kim2010multi}, and 
managing risks and events \citep{ahn2003flexible, bodendorf2005proactive, giannakis2011multi, bearzotti2012autonomous, proselkov2024financial}.

As discussed above, existing research has mainly focused on using the MAS approach to address specific functions or processes within supply chains, with less attention given to automating integrated processes.
However, achieving ASCs would require seamlessly integrating multiple processes across company boundaries to autonomously managing the three supply chain flows \citep{xu2023autonomous, nitsche2023impact}.
Additionally, existing work has relatively low technical readiness \citep{karnouskos2016key} and the implementation of more complex problems are rare \citep{nitsche2023impact}. 
This paper thus aims to address these gaps. 
We study on a more complex scenario involving integrated processes and multiple tiers of supply chain stakeholders, automating its entire physical flows and their associated information flows using the MAS approach. 
Moreover, we adopt a software engineering methodology for system design and analysis, aiming to achieve higher technical readiness and reduce the divide between academic exploration and industrial adoption.

\section{Agent-based Autonomous Supply Chain}\label{sec:a2sc}
The adoption of the Industry 4.0 technologies \citep{wef2017impact} for transforming supply chains has gained significant attention \citep{blueyonder2020autonomous, calatayud2019self}, with existing academic and industrial efforts mainly focusing on using AI, IoT, and advanced robotics for enhancing specific functions, such as demand forecasting and pallet picking, rather than automating integrated processes.

Automated functions are ``building blocks'' of an integrated automated process, which, in turn, collectively form an entire automated system.
An ASC involves numerous automated processes, autonomously managing the three flows.
Automating all specific functions needs both technological and managerial efforts \citep{nitsche2023impact}, which is beyond of the scope of this paper. 
Therefore, this paper focuses on integrating these ``already-automated'' functions and facilitating entire processes. 
Effective coordination mechanisms, both within individual business entities and across them, are crucial for achieving a seamless ASC.
This section describes the MAS approach, with a particular focus on the methodology for analysing and designing A2SC systems.

\subsection{Agents and Multi-Agent System}
Agents are goal-directed computation entities that act autonomously based on their designated roles, while also communicating and coordinating with other agents as needed \citep{wooldridge1995intelligent, fox2001agent}.
An MAS consist of a collection of such agents, interacting to address complex problems that exceed the capacities or knowledge of any individual agent. 
The MAS approach is therefore well-suited for addressing problems involving:
1) multiple (semi)-autonomous entities;
2) distributed knowledge among many entities;
3) the need for entities to coordinate to accomplish specific tasks. 
Supply chains, consisting of a set of distributed business entities, exemplify such problems. 
Consequently, the MAS approach is an ideal choice for constructing ASC systems by coordinating automated functions and processes. 
The resulting ASCs are referred to as A2SCs.

\subsection{Multi-Agent System Approach}\label{sec:mas_approach}
\begin{figure}[t]
    \centering
    \includegraphics[width=0.625\textwidth]{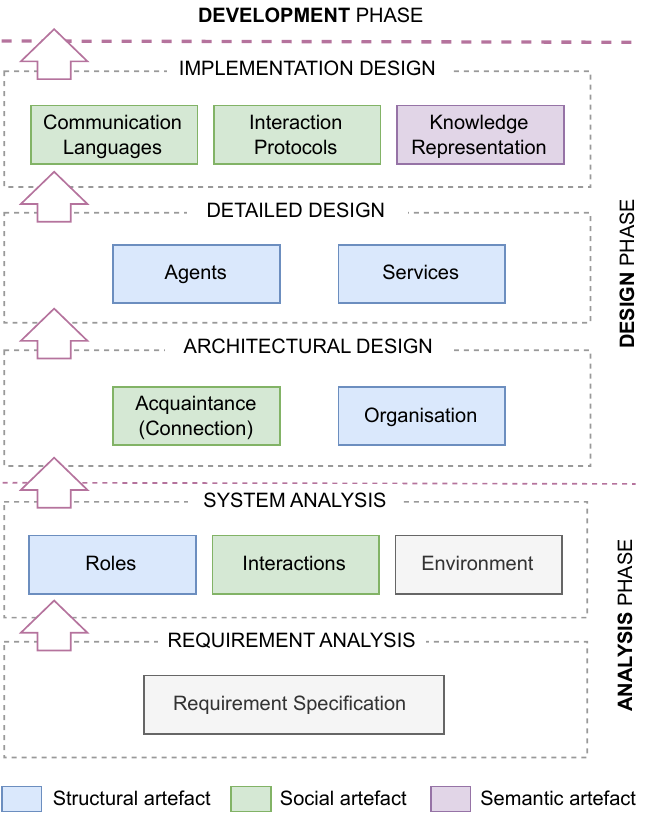}
    \caption{Methodology for the analysis and design of A2SC systems.}
    \label{fig:asc_design_methodology}
\end{figure}
Various methodologies have been proposed to analyse and design multi-agent systems, and the Gaia methodology \citep{wooldridge2000gaia, zambonelli2003developing} has gained recognition for its simplicity and effectiveness.
However, Gaia has certain limitations; it lacks completeness in abstraction and has no capacity to identify organisational rules and model environment. 
Consequently, Gaia could be only used for {\it closed} communities with cooperative agents.
To address these limitations, \citet{juan2002roadmap} introduced ROADMAP, an extension of Gaia that incorporates additional models into its framework.
Additionally, \citet{zambonelli2003developing} extended Gaia by using organisational abstractions, resulting in an extended Gaia approach capable of analysing and designing complex {\it open} systems with self-interested agents.
In contrast to ROADMAP, which is considered preliminary, this extended Gaia is comprehensive and provides coherent integration with the original Gaia approach.
We therefore chose to employ Gaia-based methodologies for the analysis and design of ASC systems, refining, customising, and adapting them to the specific context of supply chains, which are open systems and often involve self-interested agents.

The main adaptions includes: 1) we extract the models and elements from the Gaia methodology and its variants that are most relevant to the A2SC design and analysis; 2) we add implementation stage after the detailed design stage of the Gaia process, which has been specifically introduced in the Gaia methodology; and 3) we categorise the artefacts in the methodology into three groups: structural, social, and semantic, to better guiding the analysis and design of the static structure and dynamics of the A2SCs.
\figureautorefname\ref{fig:asc_design_methodology} illustrates the adapted methodology for analysing and designing A2SCs, with artefacts of different categories highlighted by blue, light green, and light violet, respectively.
This adapted methodology comprises {\it five} key stages, one more stage than the Gaia methodology, each involving specific primary tasks and output models, as outlined below:
\begin{enumerate}[label=(\arabic*), nosep]
    \item {\it Requirement Analysis}: 
    This stage determine the expectations of stakeholders for a new system under development or modification.
    The primary output artefacts are {\it requirement specifications}. 

    \item {\it System Analysis}: 
    In this stage, an understanding of the system and its structure is developed.
    It includes the identification of primary roles, the environment in which they operate, and their interactions with other roles within this system. 
    These roles serve as abstract constructs to conceptualise and understand the system.
    The primary outputs of this stage include {\it roles} model, {\it interaction} models, and an {\it environment} model. 
    
    \item {\it Architectural Design}:
    This stage defines the architectural aspects of these roles, including their acquaintance relation, organisational rules and structure.

    \item {\it Detailed Design}:
    Detailed definition of {\it agent} models and {\it services} models are developed in this stage based on the roles, their interactions, and their organisation identified in the previous stages.
    
    \item {\it Implementation Design}:
    While the previous stages mainly focus on analysing and designing the system's structures without depending on specific implementation details, this stage is dedicated to designing elements that facilitate the system's dynamics, particularly the flow of information through messaging. 
    It involves three critical aspects related to agent interaction: knowledge representation, interaction protocols, and communication languages. 
    Although these aspects do have relevance to implementation, they are included in this model as an independent design stage because the coherence of an MAS relies heavily on effective messaging. 
    Other concrete system implementation aspects are excluded from this stage.
\end{enumerate}

The methodology outlined above describes the two main phases (analysis and design phrases) that come before the development phase of implementing an ASC system.
Each of these phrases comprises several stages, with each stage containing a set of key artefacts, as illustrated in \figureautorefname\ref{fig:asc_design_methodology}.
Structural artefacts, represented in light blue in \figureautorefname\ref{fig:asc_design_methodology}, are those models related to the composition and organisational aspects of an MAS. 
Social artefacts, highlighted in light green, pertain to the interactions and relationships among agents.
Semantic artefacts, indicated in light violet, concern the understanding of the interaction content.
These artefacts collectively contribute to the analysis and design of an ASC system, providing conceptual guidelines for subsequent development phrase. 
In the following paragraphs, we delve into these artefacts by reorganising them into the four concrete issues when employing an MAS approach \citep{smith1981frameworks, genesereth1997agent, fox2001agent}.

\paragraph{Decomposition and Organisation}
In an MAS, as described in \citep{wooldridge1995intelligent}, multiple agents collaborate to achieve collective behaviours.
These individual agents typically lack sufficient information or capability to independently solve the entire problem. 
Therefore, effective collaboration among these agents is essential to ensure the system operates coherently. 
To facilitate this collaboration, it is crucial to thoroughly decompose the problem and identify specific roles that agents should play in addressing the problem. 
Agents can be assigned one or more specific roles, defining their expected behaviours (or services) within the system.
Additionally, it is important to organise these agents into a specific organisational structure that defines the topology of the interaction patterns and the control regime for activities within this organisation. 
The structural artefacts, which includes roles, agents and their services, and organisation as illustrated in \figureautorefname\ref{fig:asc_design_methodology}, are the main models for representing the composition of an MAS.
While this issue tackles the static structural aspects of an MAS, the remaining three issues focus on its dynamics, including interactions and elements that facilitate communication.

\paragraph{Acquaintance and Interaction}
Collaboration among agents is achieved through interaction. 
The design of agent interaction involves two key activities: 1) identification of the {\it acquaintance} relationship between agents, which determines whether two agents have a connection, i.e., communication paths, and 2) elaboration on the interaction process between these connected agents. 
The primary output of these two activities are social artefacts: acquaintance (or connection) models and interaction models. 
These models help identify any potential communication bottlenecks, ensuring the agents are loosely coupled. 
Well-designed interaction mechanisms can lower communication overhead, resulting in reduced data transmission and enhanced responsiveness.

\paragraph{Interaction Protocol and Communication Language}
While the first two issues define the structure of an MAS, this issue deals with dynamics, specifically the flow of messages along the communication channels between its constituent agents. 
This includes establishing protocols that regulate interaction process and defining communication languages for representing messages. 
Some representative works in these areas include 
the Contract-Net protocol \citep{davis1983negotiation},
and the English auction protocol \citep{fipa2001english}, and 
KQML \citep{finin1994kqml}, and 
FIPA-ACL \citep{fipa2002agent, poslad2007specifying}.

\paragraph{Vocabulary and Knowledge Representation}
With the first three issues addressed, the flow of messages is facilitated; however, message comprehension remains unaddressed.
To ensure that agents understand the content of messages, they must use a standardised vocabulary with teams that consistently refer to objects, functions, and relations relevant to a particular application. 
This consistent vocabulary, along with a commonly understood knowledge representation, constitute the fourth issue that requires attention when applying an MAS approach.
A vocabulary may encompass multiple {\it ontologies} used to describe the shared body of knowledge within a specific domain.

These four issues are essential when adopting an MAS approach.  
The first two issues pertain to the structural aspects (composition and organisation) of an MAS, whereas the last two relate to its dynamics, specifically, interaction and communication.
Commonly communication languages and interaction protocols allow agents to freely join or leave the MAS, enabling an open, dynamical computational environment. 
These four issues are a concrete refinement of the methodology illustrated in \figureautorefname\ref{fig:asc_design_methodology}.
Despite the differences between supply chains and traditional MAS scenarios, these issues provide valuable guidance for the analysis and design of a A2SC system.

\subsection{A2SC Analysis and Design}\label{sec:design_a2sc}
An ASC, as described in \citet{xu2023autonomous}, is an self-operating supply chain comprised of a set of structural entities. 
Each of these entities may represent individual business entities in a supply chain and can manage multiple automated processes.
These structural entities are crucial components in the formation of an ASC and are responsible for managing the three flows within the supply chain. 
Efficient operation of an ASC requires effective coordination among these structural entities. 
However, their coordination is challenging due to their inherently self-interested, distributed, heterogeneous, yet cooperative nature. 
Fortunately, the MAS approach is well-suited for building open systems that involve both cooperative and self-interested entities.

Considering the broad spectrum of automated processes involved in an ASC, often utilising technologies from various domains, delving into their specific designs of these processes is neither necessary nor feasible within the scope of this article.
Instead, we assume these automated processes as readily available building blocks for the construction of an ASC. 
Our main focus then shifts to integrating these processes and their corresponding structural entities, enabling seamless and autonomous collaboration.
This section discusses the concrete A2SC analysis and design using the MAS methodology presented in the previous section.

\subsubsection{Decomposing ASC into Agents}
In distributed problem-solving scenarios, such as distributed sensing and air traffic control, agents often share identical owners and are explicitly designed to work cooperatively towards a given goal.
In such cases, problem decomposition typically involves organising roles and agents based on functionality or geographical location.
However, in an ASC scenario, agents may represent different business entities that collaborate for mutual benefits, but they may also compete with others and thus have self-interests to consider. 
Moreover, ASCs are considered {\it open} systems that are collectively designed by stakeholders, in which stakeholders are free to join or leave this system.

In the context of A2SC, problem decomposition must take into account organisational boundaries.
According to \citet{xu2023autonomous}, structural entities act as bridge points for business entities, allowing them to connect with external peers while managing their internal entities within a supply chain.
Decomposing the supply chain into a set of structural entities, where each may represent multiple roles and potentially be represented by an agent, aligns with the natural structure of the supply chain.

Despite operating at a higher level of abstraction, this decomposition approach is consistent to some extent with existing literature.
Previous studies, such as those by \citet{swaminathan1998modeling, fox2001agent, akanle2008agent}, identified two main categories of agents: structural and functional, based on their organisational and functional roles. 
Structural agents represent the organisational entities involved in a supply chain, including stakeholders such as suppliers, retailers, third-party logistics (3PL) providers, distribution centres, and manufacturing plants. 
Functional agents manage specific tasks in a supply chain, such as supplier selection, order placement, and logistics arrangement.
A structural entity represent an conceptual organisation, thus it representative agent may oversee multiple structural and functional agents.

\subsubsection{Designing Interactions and Messages}
These decomposed agents collectively manage the operations of a supply chain through interactions.
Agent interaction defines the potential interdependencies among agents in an MAS, thus influencing the MAS's topology. 
Therefore, interactions among agents must be carefully designed to create a loosely-coupled, scalable MAS which agents focus more on computation rather than communication. 
In A2SCs, the management of the three flows --- product, information, and financial --- involves agents interacting with others to facilitate the inbound and outbound of movement of these flows. 
Developing appropriate interactions is crucial to ensuring the efficient coordination of these flows through a network of structural entity agents. 
During interactions, agents may exhibit two types of behaviours: {\it proactive} and {\it reactive}. 
Proactive behaviours involve agents initiating interaction to pursue their objectives, such as connecting other agents to request price data. 
Conversely, reactive behaviours respond to external events, such as handling incoming requests when receiving goods. 
Furthermore, interactions should be governed by protocols, which are the commonly recognised procedures or system of rules. 
These protocols manage various interaction processes, each regulating agents' dialogue with others in different scenarios, such as supplier selection, procurement price negotiation, and advanced shipment notification.

Agents interact through messages; the entire A2SC system functions through the exchange of these messages.
To function efficiently, these messages must be fully comprehended both syntactically and semantically by all participating agents.
Metaphorically, agents in an A2SC community must ``speak'' a common language to enable clear and unambiguous communication.
This common language should exhibit consistency in below critical aspects: 
\begin{itemize}[nosep]
    \item {\it Message Structure}: the way in which a message is organised or structured to convey information. 
    \item {\it Representation of Message Content}: the method used to represent the actual data contained in messages. 
    \item {\it Ontologies}: the shared vocabularies that describe the knowledge as a set of concepts within specific domains of a supply chain and the relationship  between them.
\end{itemize}

These three aspects collectively standardise the messages in terms of their structure, content representation, and semantics. 
To achieve universally understood messages, they should be explicitly specified using common languages.
Information flow among agents, including quotations, purchase orders, delivery status, and invoices, relies on effective messaging. 
Consequently, careful considerations on interaction protocols, languages, and message representation is crucial for the effective implementation of an A2SC.

\section{A Case Study: an Autonomous Meat Supply Chain}
\label{sec:a2sc_implementation}
\begin{figure}[t]
    \centering
    \includegraphics[width=\textwidth]{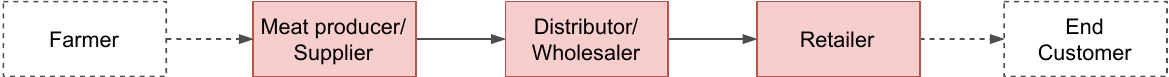}
    \caption{Illustration of a {\it simplified} meat supply chain. 
             The middle parts of this supply chain are the focus of our prototype.}
    \label{fig:meat_sc}
\end{figure}
The previous section introduces the A2SC and describes a methodology for its analysis and design.
This section details a case study of A2SC implementation to illustrate the development of an ASC prototype using an MAS approach.
We selected the meat supply chain --- a particular perishable foods supply chain --- as the case for study. 
Perishable foods, unlike other goods, are prone to spoilage due to inadequate storage conditions.
It is crucial to maintain proper storage conditions and minimise transit times to ensure the successful delivery of perishable foods.
Monitoring the environmental conditions of trucks transporting perishable foods and swiftly addressing unforeseen events, such as traffic congestion, is of paramount importance.

In this case, we focus on an integrated process --- automated meat procurement --- specifically implementing an autonomous process for procuring meat and transporting meat from producers to end customers.
This procurement process involves several key stakeholders across the meat supply chain. 
However, it is important to note that our representation of meat supply chain is simplified and simulated (\figureautorefname\ref{fig:meat_sc}). 
It excludes minor participants, such as secondary processors, which thus may not exactly mirror the complexity of real-world meat supply chain operations.
\figureautorefname\ref{fig:meat_sc} presents an illustration of this supply chain. 
Our description mainly focuses on the central process highlighted in red.

Furthermore, this prototype includes logistics and 3PLs since they are responsible for transportation, which are essential to perishable food supply chains. 
The core procurement process involving suppliers, wholesalers, and retailers is sufficient for illustrating an initial A2SC implementation; and including additional entities would lead to unnecessary redundancy.

\subsection{Scenario: the Cambridge Meat Company}\label{sec:scenario}
The prototype is based upon the following scenario:
\begin{displayquote}
    The Cambridge Meat Company (CMC), a hypothetical company used for illustration purposes, specialises in wholesale meat procurement and supplies local restaurants. 
    The CMC has established long-term contracts with its meat suppliers and aims to automate its wholesale and procurement process by incorporating disruptive information and communications technologies. 
    Specifically, the company seeks automation in four key areas: 
    \begin{enumerate}[label=\arabic*), nosep]
        \item Selection of bids from suppliers;
        \item Monitoring of the logistics process;
        \item Adaptation to unforeseen events;
        \item Evaluation of the quality of both the logistics service and the supplied products.
    \end{enumerate}
\end{displayquote}

From this scenario, we can further analyse the requirements for the prototype, aiming to develop an integrated automated process for meat procurement and wholesale.
A crucial requirement for this prototype is to ensure the smooth movement of physical flow and its corresponding information flow. 
Given that modern financial transactions primarily rely on digital information, essentially constituting a specialised from of information flow, this prototype does not encompass financial flow considerations. 
Additionally, this prototype take into account inventory management, continuously updating it as goods are received and dispatched. 
The following sections first extract requirements and then detail the design, implementation, and showcase of the A2SC prototype. 

\begin{figure}[t]
    \centering
    \includegraphics[width=\textwidth]{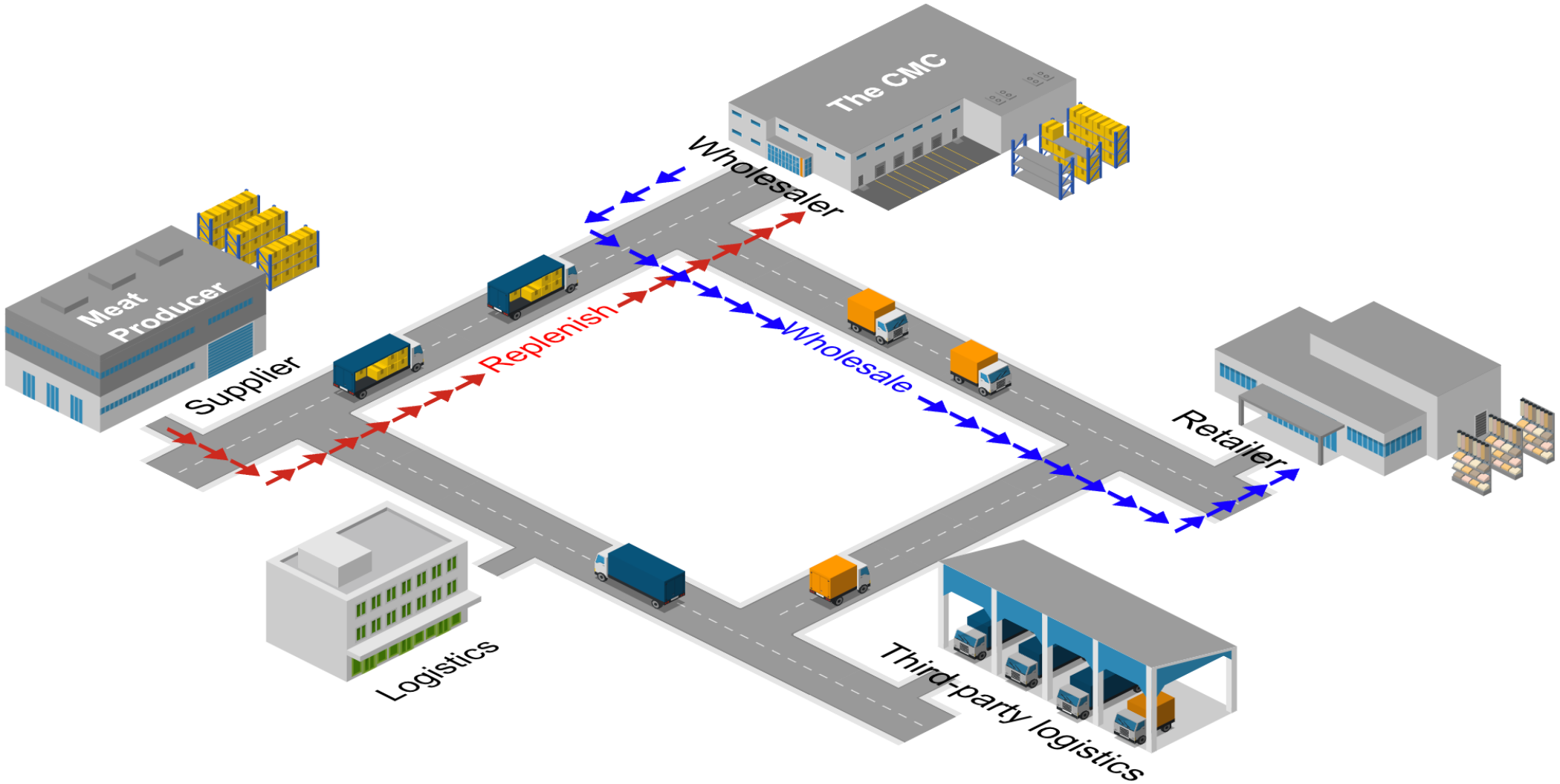}
    \caption{Illustration of the CMC scenario.}
    \label{fig:cmc_process}
\end{figure}

\subsection{Process Design}\label{sec:process_design}
\begin{figure}[t]
     \centering
     \begin{subfigure}[b]{0.485\textwidth}
         \centering
         \includegraphics[width=\textwidth]{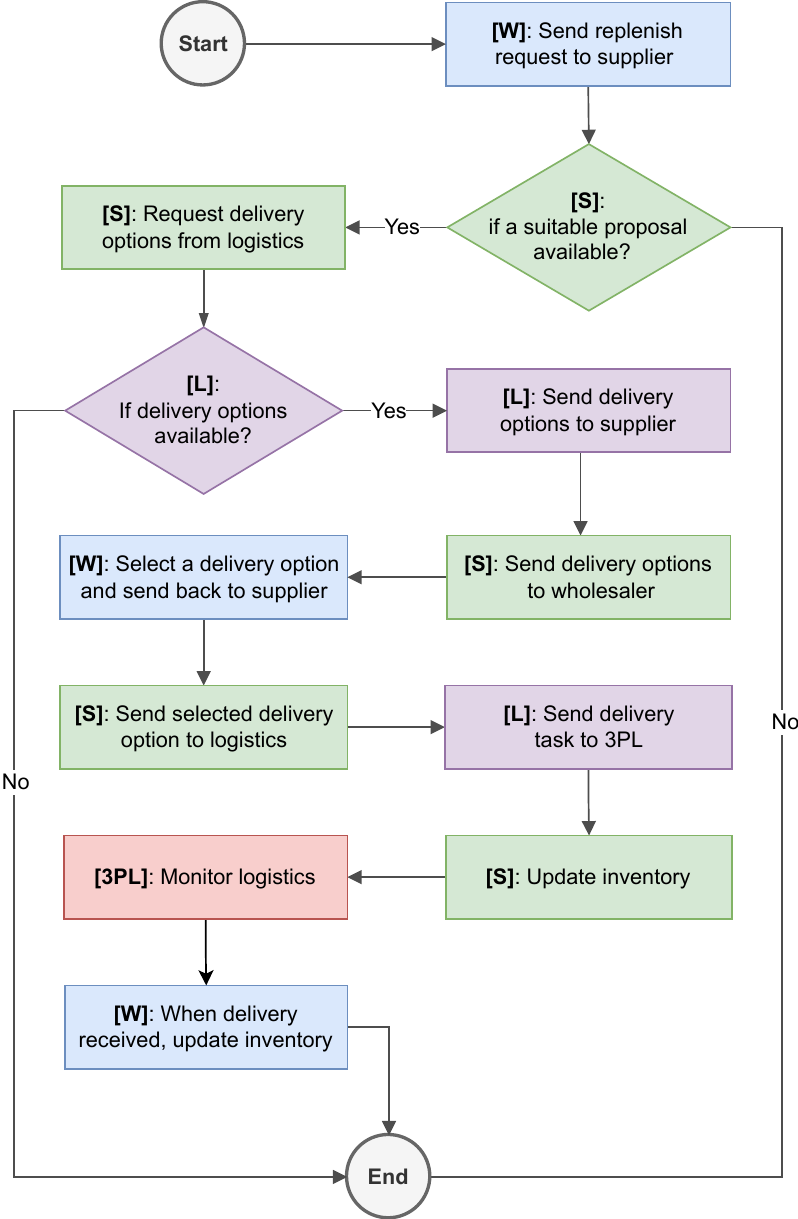}
         \caption{Replenishment}
         \label{fig:replenish_process}
     \end{subfigure}
     \hfill
     \begin{subfigure}[b]{0.485\textwidth}
         \centering
         \includegraphics[width=\textwidth]{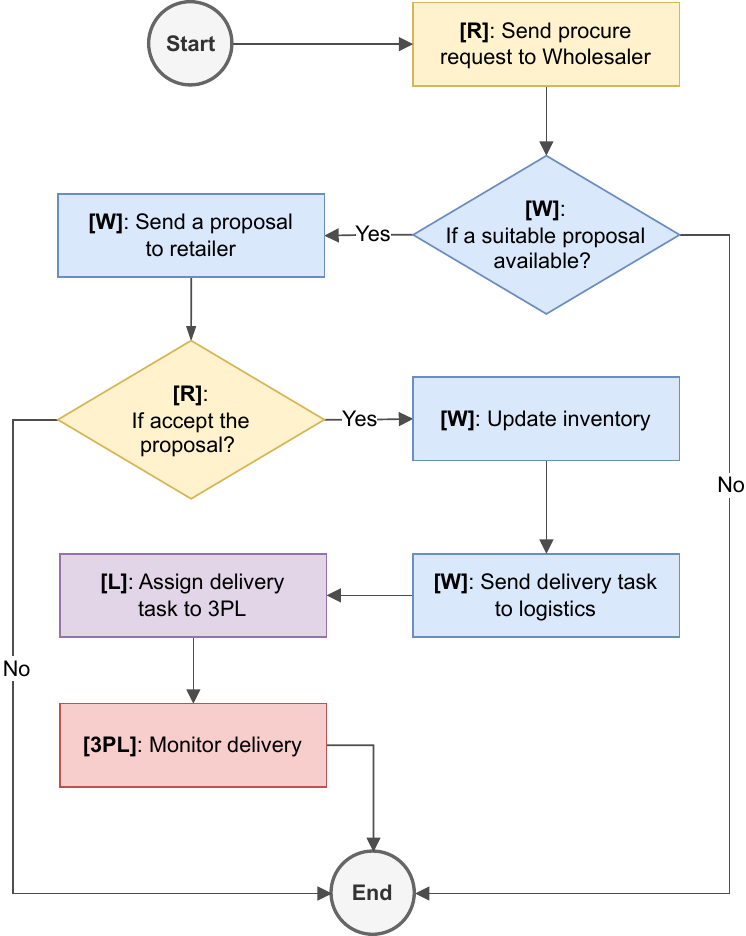}
         \caption{Wholesale}
         \label{fig:wholesale_process}
     \end{subfigure}
     \caption{Flowcharts illustrating the two main processes: replenishment and wholesale. 
             Uppercase letters denote supply chain stakeholders: `S' for supplier, `W' for wholesaler, `R' for retailer, `L' for logistics, and `3PL' for third-party logistics provider.}
     \label{fig:two_processes}
\end{figure}

This prototype includes two primary processes: 
{\it replenishment} (the CMC procures meat from its suppliers to replenish its inventory) and
{\it wholesale} (the retailer buys meat from the CMC).
\figureautorefname\ref{fig:cmc_process} presents a diagram of the refined CMC scenario, with the two processes highlighted by the flow of blue and red arrows.
In addition to the three main stakeholders (supplier, wholesaler, and retailer), logistics and 3PL, which provide logistics service, are included. 
Logistics service and inventory management, crucial for the flow of goods (meat), are considered in both processes.
Both processes include logistics services, with sellers (supplier and wholesaler) responsible for delivery arrangements. 
\figureautorefname\ref{fig:two_processes} illustrates the flowcharts of these two processes, highlighting major steps. 
Different supply chain stakeholders are denoted by capital letters in brackets and are represented using different colours.
The replenishment process (see \figureautorefname\ref{fig:replenish_process}) provides customers (wholesalers) with multiple delivery options, allowing them their preferred delivery service. 
However, the wholesale process (see \figureautorefname\ref{fig:wholesale_process}) uses a predetermined delivery option selected by the wholesaler for their customers (retailers), making it less complex.
Both processes involves multiple decision-makings points (denoted by diamonds), such as proposal evaluation, acceptance or rejection of proposals, and delivery option selection. 
Inventory-related functions, such as automatic inventory updates and replenishment, are integrated into both processes.
Additionally, both processes incorporate logistics monitoring, which tracks real-time locations and ambient conditions of goods throughout delivery.

\subsection{Roles, Agents, and Services}\label{sec:agents}
\begin{figure}[t]
    \centering
    \includegraphics[width=0.90\textwidth]{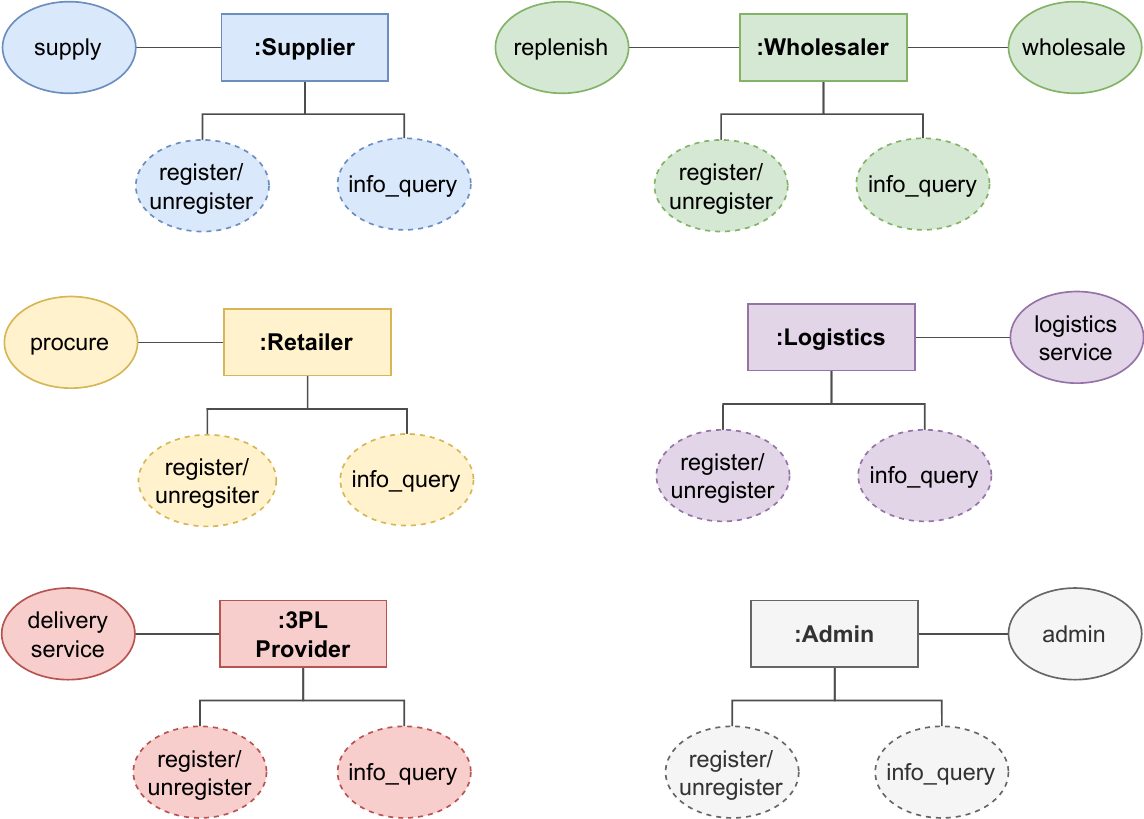}
    \caption{Agents and heir main} services.
    \label{fig:agent_services}
\end{figure}
From the scenario described in \autoref{sec:scenario} and its subsequent refinement processes (see \figuresautorefname\ref{fig:cmc_process} and \ref{fig:two_processes}), we identified five distinct types of roles: suppliers, wholesalers, retailers, logistics providers, and 3PL provider, and each with their corresponding primary services, which are a coherent blocks of functions that roles are expected to accomplish. 
Each of these roles can be represented by individual agents, serving as the {\it structural entities} within the scope of their respective organisations.
These agents and their main services are shown in \figureautorefname\ref{fig:agent_services}.
Each agent is detailed as follows:
\begin{enumerate}[label=(\arabic*), nosep]
    \item {\it Wholesaler Agent}: 
    This agent represent the central role of the wholesaler (the CMC) in this scenario.
    Its main services involve managing the procurement of meat from supplier agents,  wholesaling meat to retailer agents, and overseeing inventory during these processes.
    It also coordinates with logistics agents to arrange delivery service for customers, functioning as a central hub connecting upstream and downstream supply chain agents, facilitating entire flows.

    \item{\it Supplier Agent}:
    This agent supplies meat to wholesaler agents on behalf of a supplier. 
    Like wholesaler agents, the supplier agent outsources delivery services to logistics agents, necessitating coordination with both wholesale agents and logistics agents to fulfill purchase orders.

    \item{\it Retailer Agent}:
    Representing local retailers, such as restaurants or local stores, this agent procures meat from wholesaler agents and interacts with them to manage procurement on behalf of the retailer.

    \item {\it Logistics Agent}:
    This agent, representing a logistics company, offers logistics services to supplier agents and wholesaler agents in the scenario. 
    Logistics companies manage overall logistics service but outsource fulfilment service to 3PL providers. 
    Logistics agents coordinate with suppliers, wholesalers, and 3PL providers for delivery arrangements.

    \item {\it 3PL Agent}:
    Acting on behalf of a 3PL provider, this agent provides delivery services to logistics companies, fulfilling delivery orders assigned by a logistics agent. 
    It monitors the entire meat transportation process, supplying real-time data on the truck's geolocation and ambient conditions for logistics companies to use, such as in fulfilment service evaluation.
\end{enumerate}

In addition to the main services discussed above (represented by solid-lined ellipses in \figureautorefname\ref{fig:agent_services}), these agents provide information query services to other agents and include functions for (un)registration.
Information query service enable agents to request specific information from other agents. 
They can register their services for discovery or unregister existing services, removing them from the system and making them unavailable for discovery by other agents.

These five types of agents are the main stakeholders --- structural entities --- in this meat supply chain, and their interactions form the dynamics of this supply chain. 
Additionally, we introduced a functional agent, the admin agent, responsible for managing agent-level administration services, including providing yellow page services.
The following sections describe the organisation and interactions among these agents.

\subsection{Connection Mechanism}\label{sec:connection_mechanism}
\begin{figure}[t]
     \centering
     \begin{subfigure}[b]{0.485\textwidth}
         \centering
         \includegraphics[width=\textwidth]{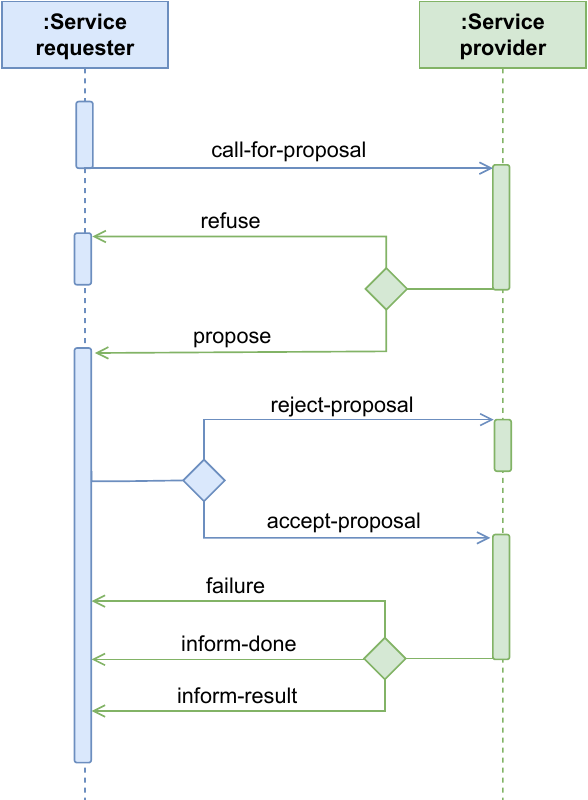}
         \caption{Contract net protocol}
         \label{fig:contract_net_protocol}
     \end{subfigure}
     \hfill
     \begin{subfigure}[b]{0.485\textwidth}
         \centering
         \includegraphics[width=\textwidth]{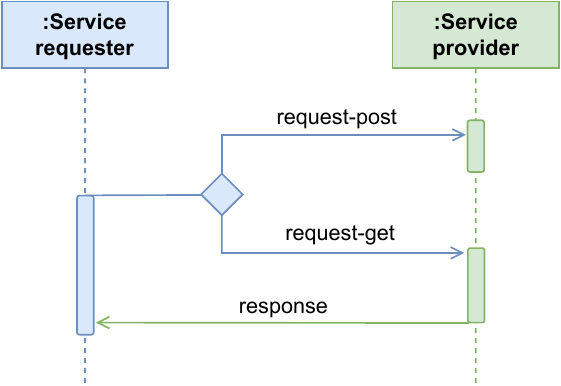}
         \caption{HTTP protocol}
         \label{fig:http_protocol}
     \end{subfigure}
     \caption{Interaction diagrams for the contract net protocol and the HTTP protocol. 
              Note: The left is the FIPA contract net interaction protocol, which is a slight adaption of the original contract net protocol \citep{smith1981frameworks}.}
     \label{fig:interaction_protocols}
\end{figure}
Prior to organising distributed agents, it is crucial to enable connections among them.
We first introduce connection mechanisms to facilitate agent organisation and interactions.
In an MAS, agents offer services to others or request services from others, matching service requester with suitable service providers. 
This is known as the {\it connection} problem \citep{davis1983negotiation, ameri2013multi} in distributed problem solving.

In this prototype, we devised two agent connection mechanisms.
The first is the Open Economic Framework (OEF)-based connection, where two agents are linked through a third-party service, the OEF. 
In this indirect or mediated connection, a service requester queries the OEF, which responds with a list of potential service providers. 
The requester can then choose one or more providers from the response for further connections. 
The second mechanism is the HTTP-based connection, where two agents connect directly through the HTTP protocol (see \autoref{sec:interaction}).
In this connection, a service requester sends an HTTP request to a service provider via its URL and receives an HTTP response.
This connection is suitable for scenarios where both agents are already acquainted and can contact each other directly without an intermediary.

\subsection{Agent Organisation}\label{sec:agent_organisation}
\begin{figure}
    \centering
    \includegraphics[width=0.90\textwidth]{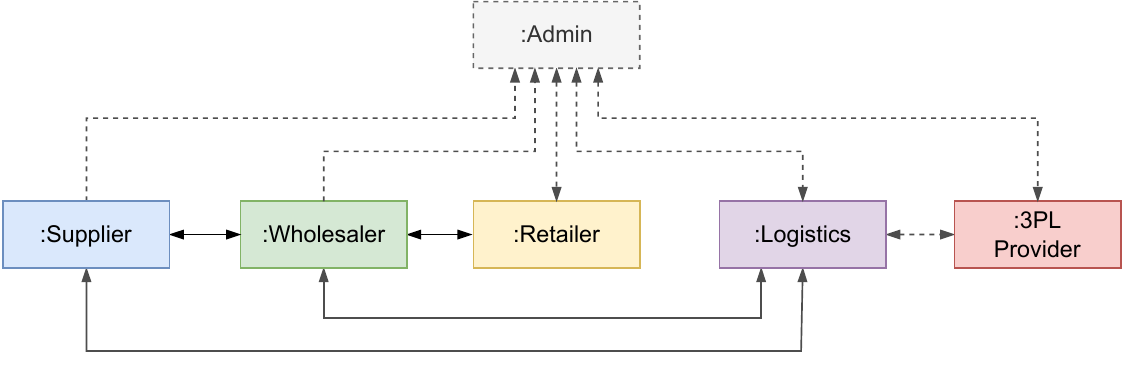}
    \caption{Agent organisation, with OEF and HTTP-based connections denoted by solid and dashed lines, respectively.}
    \label{fig:agent_organisation}
\end{figure}
Connection mechanisms facilitate agent connections, and for coherence, these agents should be organised in a specific structure --- agent organisation --- which defines the way of agent communications and connections \citep{horling2004survey, dorri2018multi}.
As described in \autoref{sec:agents}, this prototype includes five types of structural agents and a functional agent. 
These functional agents are self-interested but may collaborate for a short-term goal such as order fulfilment or build a long-term partnership for outsourcing logistics services.
Intermediary agents such as blackboard, yellow page, and broker were not included in this design due to: 
a) this prototype involves a small number of agents, and 
b) the OEF is capable of providing efficient discovery and matchmaking services.
Therefore, we employ a compound organisation paradigm (see \citep{horling2004survey} for more details about this organisation), in which multiple organisation forms (such as coalition and team) are adopted for organising the relationship among agents, as illustrated in \figureautorefname\ref{fig:agent_organisation}.
For better illustration, although unnecessary for an organisational diagram, we differentiate the two connections discussed earlier between agents in this diagram, denoting them with solid and dashed lines, respectively.
These agents operate with independent control regimes, each focusing on specific stages of the three flows.
The admin agent can establish connections with all other agents via the HTTP connection,  maintaining an portal that provides an overview of the status of all agents in the system. 
The OEF connection links the supplier, wholesaler, and retailer agents, enabling negotiations for replenishment or procurement. 
The HTTP connection becomes active between these agents if trust is established.

\subsection{Agent Interactions, Language and Protocols}\label{sec:interaction}
While agents and their organisation define the A2SC system's structure, this section discusses its dynamics, focusing on interactions among agents and the language and protocols  facilitating these interactions. 
As shown in \figureautorefname\ref{fig:two_processes}, the wholesale agent interacts with supplier agents in the replenishment process and with retailers during the wholesale process. 
Both processes require logistics services, involving two additional agents: logistics and 3PL provider agents. 
These agents interact by following agreed-upon interaction protocols.

\begin{figure}[!t]
    \centering
    \includegraphics[width=\textwidth]{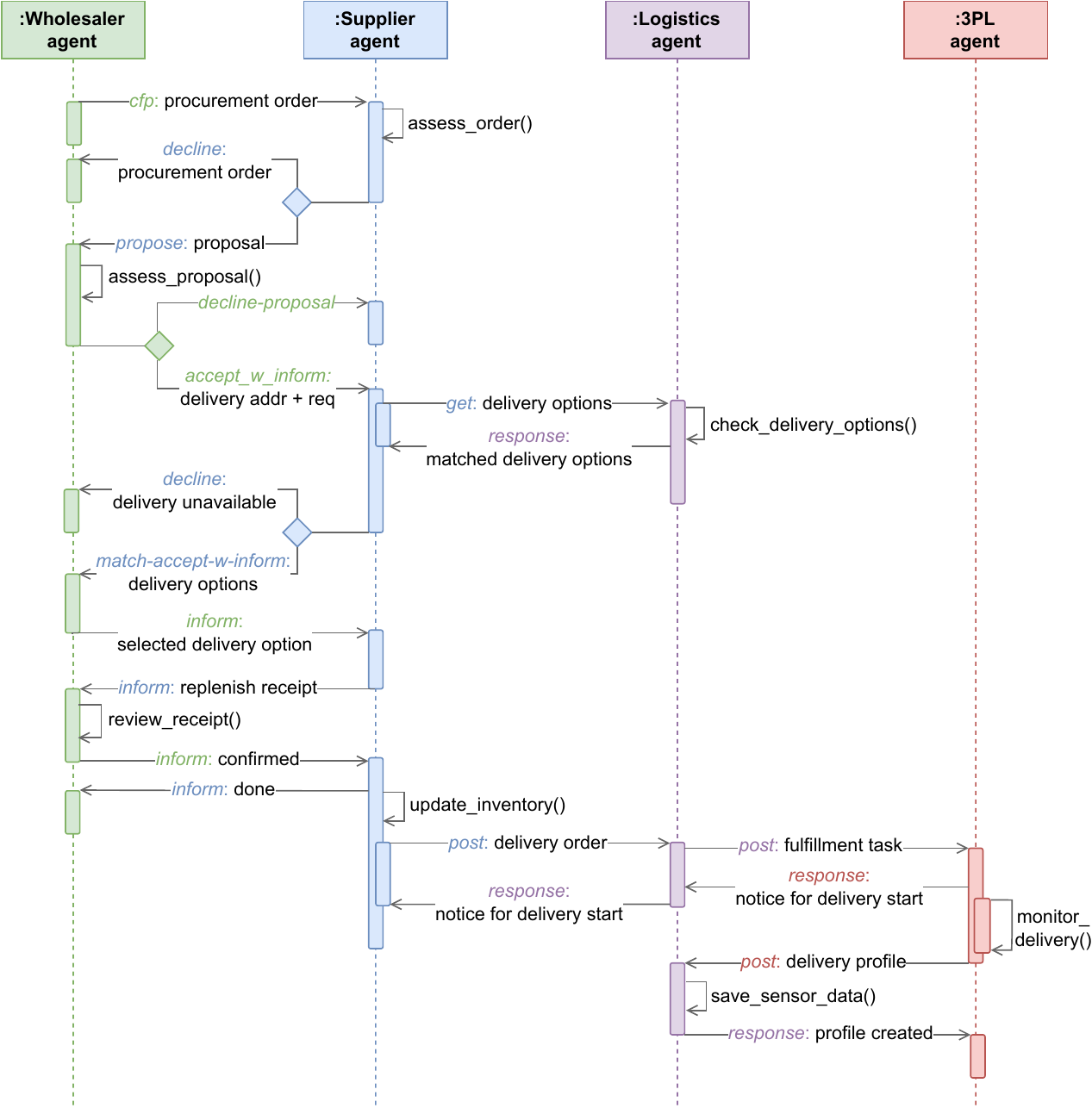}
    \caption{Agent interaction in the replenishment process.}
    \label{fig:agent_interaction_replenish}
\end{figure}
\begin{figure}[!t]
    \centering
    \includegraphics[width=\textwidth]{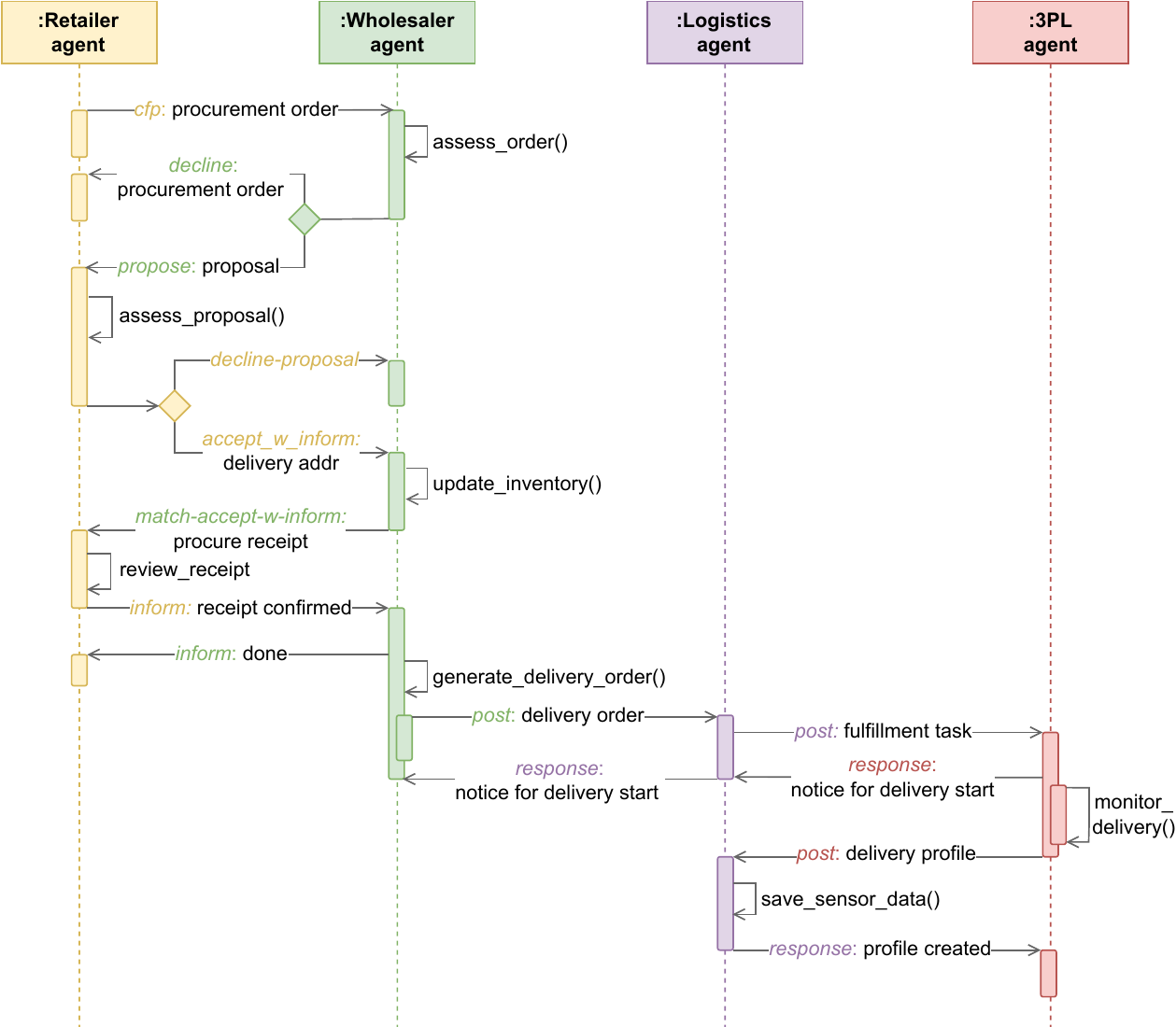}
    \caption{Agent interaction in the wholesale process.}
    \label{fig:agent_interaction_wholesale}
\end{figure}

In this implementation, we incorporate two types of interaction protocols: the contract net protocol and the HTTP protocol. 
The contract net protocol is specifically designed for scenarios where a service requester needs to select suitable service providers for executing a specific task, often involving multiple rounds of negotiation and information exchange, as illustrated in \figureautorefname\ref{fig:contract_net_protocol}. 
It is well-suited for managing complex interactions with {\it unknown} participants, such as procurement.
Conversely, the HTTP protocol is adopted for simpler interactions typically involve requiring only a single round. 
As shown in \figureautorefname\ref{fig:http_protocol}, an HTTP client (service requester) can send a {\it request} to an HTTP server (service provider) to either {\it get} data or {\it post} a task for execution.
This protocol is most suitable for interactions between known participants with established trust, such as a logistics company assigning a delivery task to their longstanding 3PL partners.
Additionally, the HTTP protocol is appropriate for quickly retrieving information such as product wholesale price, delivery rates, and current traffic conditions.

The detailed interactions among agent in the two processes are illustrated in \figuresautorefname\ref{fig:agent_interaction_replenish} and \ref{fig:agent_interaction_wholesale}, 
Both processes involve four types of agents, interacting through messages. 
Messages in these figures consist of two essential parts: performative and message body, separated by a colon. 
These parts respectively represent the communicative act (behaviour) and the content of the message.
Interaction protocols define a set of performatives used in interactions. 
For example, the contract net protocol includes performatives such as {\it cfp} (call for proposal), {\it propose}, and {\it reject-proposal}. 
The HTTP protocol includes two basic performatives: {\it request} (further categorised as get and post) and {\it response}, as shown in \figuresautorefname\ref{fig:agent_interaction_replenish} and \ref{fig:agent_interaction_wholesale}. 
A message should include metadata for unambiguous communication, such as sender, receiver, protocol, and ontology.
The agent communication language defines the message format, and in this study, we use FIPA-ACL for agent communication.

Both processes involves multiple rounds of communication, sending and receiving messages to coordinate procurement and logistics service. 
They also include decision-making support procedures, such as \texttt{assess\_order()} for evaluating incoming procurement orders, and self-calls for executing specific tasks, such as \texttt{monitor\_delivery()} for tracking delivery process.
In the replenishment process, which offers multiple delivery options, there are additional rounds of interaction with the logistics agent, and subsequently with the 3PL agent, to request information about available delivery options. 
This is depicted in the right bottom area in \figuresautorefname\ref{fig:agent_interaction_replenish}.
Although the two processes are operationally separate and independent, the wholesale process may trigger the replenishment process automatically when conditions are met. 
For instance, when the wholesaler's inventory reaches a predetermined low stock level, the replenishment process is initiated.
In this sense, the two processes are interconnected, illustrating integrated supply chain processes that facilitate the automated movement of ``goods'' from suppliers through intermediaries to customers.

\subsection{Implementation}\label{sec:implementation_a2sc}
The previous sections provide an overview of the A2SC prototype's design.
This section describes the implementation details of the prototype.

\subsubsection{Toolkit for Implementation}\label{sec:toolkit}
We used Python as the primary language for developing the backend of this prototype, while other languages such as HTML and JavaScript were used for creating frontend interfaces.
To enhance development efficiency, we employed a set of tools and frameworks detailed in \ref{appx:toolkit}), facilitating communication, interaction, and visualisation functionalities for both the backend agent network and frontend interfaces in the A2SC prototype.

\subsubsection{System Architecture}\label{sec:system_architecture}
\begin{figure}
    \centering
    \includegraphics[width=\textwidth]{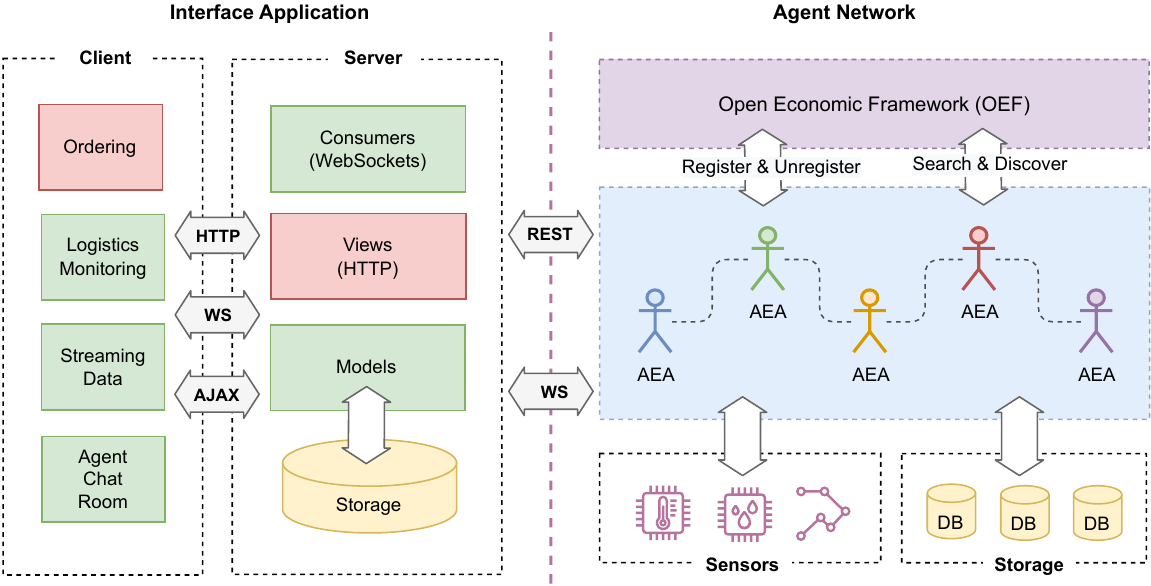}
    \caption{The system architecture of the prototype, with WS stands for the WebSocket protocol.}
    \label{fig:system_architecture}
\end{figure} 
This A2SC prototype primarily consists of two main components: an agent network and an interface web application.
We implemented a {\it microservice} architecture to integrate these components using RESTful APIs and WebSockets, as depicted in \figureautorefname\ref{fig:system_architecture}. 
The agent network is composed of a collection of autonomous economical agents (AEAs), implemented using the Fetch AEA framework as described in \ref{appx:toolkit}, with each representing distinct organisations or functions. 
These AEAs operate within an ``environment'', specifically the OEF, and collaborate cohesively to pursue their respective objectives. 
Additionally, they may also access storage services to persist data and utilise sensors to sense the environment.
The interface application, powered by Django and Channels, provides a web-based portal to the prototype. 
The interface comprises a client with a collection of UI components, which we will further described in the next section.
Additionally, it includes a server that handles HTTP and WebSockets requests through its Views and Consumers modules, respectively. 
Specifically, it handles user input and provides real-time visualisation of the agent network's state through RESTful APIs and/or the WebSocket protocol.
This architectural design enables a loose coupling between the agent network, which may be distributed over many machines, and the interfaces, allowing them to maintain independence and extensibility.

\subsubsection{Interface}\label{sec:interface}
We developed a single-page web-based interface for this prototype, based on the predefined scenario and the system architecture.
The interface consists of four main panels, each serving a specific purpose:
\begin{enumerate}[label=\arabic*), nosep]
    \item {\it Ordering}:
    This panel provides users with an interface for placing orders to purchase products. 
    
    \item {\it Logistics Monitoring}:
    This panel visualises the real-time track and trace of the vehicle fulfilling the delivery.
    
    \item {\it Streaming Data}:
    This panel displays the real-time charts of the ambient conditions of the products during transportation, including temperature, humidity, and lightness.
    
    \item {\it Agent Chat Room}:
    This panel shows the negotiation process among the agents, including message exchanges between them.
\end{enumerate}

These panels interact with the agent network using either HTTP or WebSockets. 
We use Ajax for the ordering panel and WebSockets for the other panels to dynamically update the current interface with new data from the backend systems, enhancing use experience by eliminating the need to load entire new pages.

\subsubsection{Data Preparation}\label{sec:data_preparation}
As previously described, this prototype involves real-time data collection to monitor meat delivery processes.
Instruments such as sensors and GPS trackers are required to capture the ambient condition of the meat and the location of its transportation vehicle.
This data needs to be continually collected during the delivery process. 
However, due to limited experimental condition, it is not feasible to actually drive a vehicle with necessary data collection instruments and refrigerated containers to transport meat from one place to another while establishing a wireless communication link with remote agents.
The primary objective of this study is to demonstrate the technical methodology rather than achieve a high level of technical readiness.
Employing a realistic setting for this prototype would be uneconomical and unnecessary.

We therefore used pre-collected IoT and GPS data to simulate real-time data collection during the delivery process.
These data were collected through the following steps:
\begin{enumerate}[label=\arabic*), nosep]
    \item 
    Select multiple locations in Cambridge, UK, as the locations for the suppliers, the wholesaler, and the retailer.
    
    \item 
    Install an all-in-one sensor \footnote{UbiBot WS1 was used in this experiment.} and a smartphone with a GPS tracking App (myTracks \footnote{\url{https://apps.apple.com/us/app/mytracks-the-gps-logger/id358697908}}) on a bike.
    
    \item 
    Ride the bike from one selected location to another, capturing the data about the ambient condition of the bike and its location once every five seconds over this journey.
    
    \item 
    Calibrate the collected data and save them as CSV files for subsequent use in prototype development.
    
\end{enumerate}

The collected data consists of a series of sensor (temperature, humidity, and light) and geolocation (longitude, latitude, and elevation) data points indexed in chronological order.
We utilised these data to simulate real-time data capture during the delivery monitoring.

\subsubsection{Implementation Details}\label{sec:implementation_details}
We adopted a loosely coupled architecture, in which the frontend interface and the backend agent network can be developed independently.
These two components were integrated into a system through the use of RESTful APIs and the WebSocket protocol. 
In total, we implemented {\it seven} distinct agents within the agent network, comprising one supplier, one wholesaler, two retailers, one logistics, two 3PL providers, and one admin.
These agents were individually developed using the AEA framework and can collaborate by exchanging messages by following predefined interaction protocols.
More details about implementation can be found in \ref{appx:implementation_details}.

\subsection{Showcase}\label{sec:prototype_showcase}
\begin{figure}[t]
    \centering
    \includegraphics[width=\textwidth]{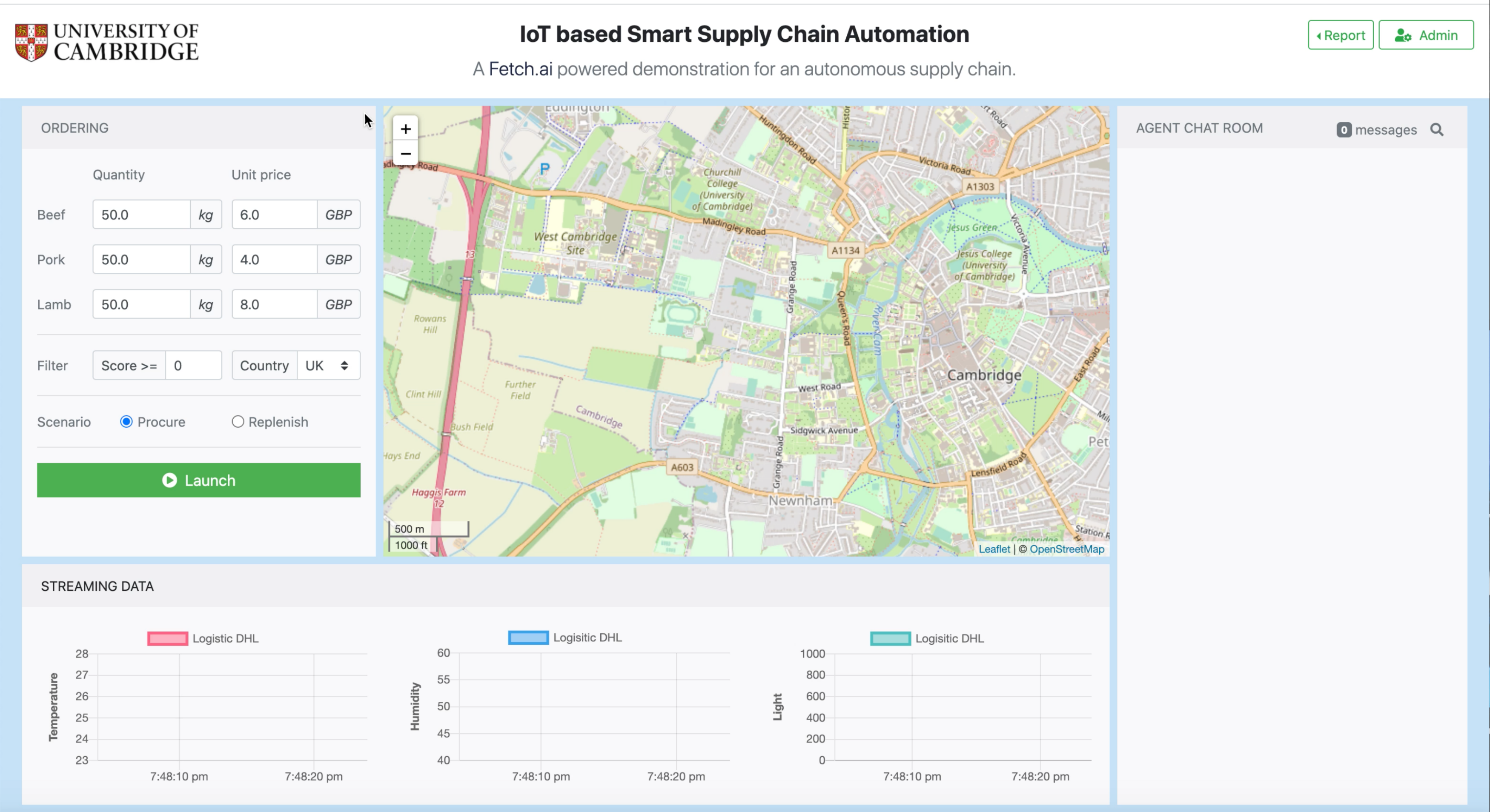}
    \caption{A screenshot of the startup interface of the prototype.
             This interface consists of four areas: the top left panel (ordering), the top middle panel (map for delivery monitoring), the right panel (agent chat room), and the bottom panel (IoT data visualisation).
             All of areas are in their initial states, containing empty content or default values.
             }
     \label{fig:interface_startup}
\end{figure}

Once the OEF, all agents, the web server, and the database are started up, the prototype is ready for running.
Upon initialisation, each agent performs a registration behaviour --- a one-shot proactive action that involves sending a message to the OEF for registering their service. 
This registration enables their discovery by other agents in the system.

The prototype can be accessed via a web browser.
\figureautorefname\ref{fig:interface_startup} shows the startup interface with four panels as described in \autoref{sec:interface}.
Initially, the interface shows no data except for default values in the ordering panel.
These default values include the default quantity and unit price of the meat for purchase, default performance and location requirements for vendors, and the default scenario.
This section showcases this A2SC prototype, focusing on the two main processes: procurement and replenishment.

At the start of execution, the wholesaler's inventory is empty. 
To prevent out of stock when the retailer procures meat from the wholesaler, we first start a replenishment process by selecting the `replenish' scenario and clicking the `Launch' button with default values.
Clicking `Launch' button triggers the replenishment behaviour of the wholesaler agent.
The wholesaler agent first establish connection with the OEF and sends a query to search suitable suppliers.
The OEF responds with a message listing supplier agents that meat the query conditions. 
This process is essentially a {\it matchmaking} step, matching a service requester with a suitable provider.
The wholesaler then engages in dialogues with discovered suppliers, negotiating with a selected supplier for purchasing products.
The logistic agent participates in this dialogue, and the supplier requests available delivery options from this agent.
Once all parties agree, the logistics agent randomly selects one of eligible 3PL agents to fulfil the delivery.
The delivery process begins; and the 3PL agent is responsible for monitoring and managing this process.

The resulting interfaces during the replenishment process are shown in \figuresautorefname\ref{fig:interface_replenish_start} and \ref{fig:interface_replenish_delivery}.
\figureautorefname\ref{fig:interface_replenish_start} shows the interface when the dialogue is completed, and the delivery is commencing. 
At this stage, no IoT data has been collected, and the charts below the map display no data.
Notifications at the top right corner, as presented in \figureautorefname\ref{fig:interface_replenish_start}, provide key progress updates, such as the completion of the dialogue and the commence of delivery.
\figureautorefname\ref{fig:interface_replenish_delivery} presents the interface during the delivery process, featuring a moving vehicle on the map with a blue-line trace, an overlaid tracking number, and a blinking red circle.
Ambient condition inside the delivery vehicle is also captured and visualised on the charts.

When the product has been successfully delivered, data collected during the delivery journey are utilised to generate a summary report using a data profiling tool. 
Access to this report is available through the `Report' dropdown button or the stacked icons at the top right corner of the map. 
The purpose of this report is to demonstrate the data's potential for evaluating the delivery service.
The corresponding interface is shown in \figureautorefname\ref{fig:interface_replenish_summary}.
The automated procurement for a retailer is similar to the wholesaler's replenishment, as described in \autoref{sec:process_design}. 
When the procurement is complete, the entire trace of the delivery vehicle is displayed on the map, as shown in \figureautorefname\ref{fig:interface_track_trace}.

The dialogues in this process take place through messages, which are displayed in the agent chat room area.
Excerpts of agent chat messages are provided in \figureautorefname\ref{fig:message_excerpts}.
These messages have four properties: sender, recipient, performative, and message body (content). 
The message body contains data (represented using JSON) associated with the message, such as procurement orders, receipts, and delivery notifications.
A more detailed demonstration, including a screen recording of this prototype, is available via this link  \footnote{\url{https://www.youtube.com/watch?v=WFcLAif-cHo}}.

\section{Discussion and Limitations}\label{sec:discussion}
The prototype provides a concrete example to demonstrate the implementation of an ASC using an MAS approach.
The prototype comprises two distinct integrated processes: replenishment and wholesale. 
Both processes can be manually initiated by a user clicking the `Launch' button after filling out an order form, or they can be set up to start automatically.
In addition to order placement, these two processes involve a set of automated functions, including supplier selection, negotiation, logistic service selection, and delivery monitoring and evaluation, inventory management. 
These functions are orchestrated into automate processes through messaging between agents and facilitate the management of the ``virtual'' product flow and its associated information flow.
While this work demonstrates a promising approach to achieving ASC, as one of the earliest technical implementations, it also comes with some limitations.

The scenario adopted in the prototype is a simplified version of real-world operations in a meat supply chain.
Its automation is mainly limited to the ordering and order fulfilment processes in a controlled simulation environment.
While targeting level 2 autonomy, which requires process automation according to \citet{xu2023autonomous}, the prototype only achieves partial process automation. 
It primarily addresses the management of the information flow while digitally representing the physical flow and disregarding the financial flow.
This simplification reduces the complexity of the ASC system analysis and design, however, it may also widen the gap between this work and realistic systems.

While this prototype successfully integrates automated functions and connects decision making points, it lacks the ability to reconfigure and adapt to disruptions, which is essential for ASCs to achieve resilience. 
However, the proposed approach lays the foundation to facilitate this capability by enhancing agents with more advanced decision making skills.
Realising this capability will be one of our primary future endeavours.
Additionally, crucial SCM functions such as planning are absent, hindering the prototype from reaching higher autonomy levels.

This prototype involves around {\it ten} stakeholders, which is significantly fewer than what is typically encountered in real-world supply chain scenarios.
Managing larger number of agents may introduce complexities in architectural scalability, protocol design, and managerial challenges etc.
In real world applications, ASCs must address these complexities to create an open, dynamic, and adaptive society of autonomous agents that can self-adjust to accommodate changes and uncertainties.

Moreover, this prototype may not fully realise a complete ASC automated processes for the three flows. 
Instead, it represents an experimental proof of concept, implementing the MISSI model, a conceptual model described in \citet{xu2023autonomous}.
This prototype mainly focuses on the two central layers of the MISSI model: interconnection and integration.
It adopts the MAS approach as the overarching architecture, facilitating the interconnection and integration of decentralised business entities across the supply chain, thereby creating a cohesive system. 
The coherence of this system relies on interactions through messaging between participating agents.
To achieve effective interactions, three key aspects need attention:
\begin{itemize}[nosep]
    \item {\it Representation}: 
    This aspect involves formally describing the current status or knowledge of an issue, request, or response.
    It pertains to how data, such as bids or events, is structured and made machine-readable.
    
    \item {\it Decision-making}: 
    This aspect concerns the evaluation of incoming requests and the process of making appropriate responses, which involves defining rules and criteria for determining the best course of actions in response to various issues. 
    
    \item {\it Social}: 
    This aspect focuses on establishing shared standards of acceptable behaviour, protocols, and languages that facilitate smooth interactions. 
    It involves defining the norms and conventions that govern agent communication and collaboration within the system.
\end{itemize}

In this prototype, we have addressed these three aspects --- representation, decision-making, and social interactions. 
However, our implementation is relatively basic and has limitations.
For representation, we focused on how agents represent their services and certain data objects, such as purchase orders and receipts.
Decision-making is rule-based and relatively straightforward. 
For instance, a wholesaler's strategy for order acceptance is based on a simple rule: accepts the order if it has sufficient stock to supply and the purchase price is higher than expected; otherwise, it rejects the order. 
We employed the contract-net protocol for regulating interactions and task coordination, 
However, this protocol alone may not be adequate for handling complex real-world interactions.
Socially, the prototype lacks formal languages for knowledge representation and social protocols. 
We underscore the need for well-designed formal or natural languages for knowledge representation and social protocols.

According to the {\it autonomy manifold} presented in \citet{xu2023autonomous}), this prototype falls within the automation-skewed region.
However, both the intelligence and automation levels of the prototype are still relatively low.
Its intelligence dimension, primarily focusing on decision-making capabilities, is basic, relying on simple rules that consider only a few main factors.
To address real-world supply chain challenges, agents need more sophisticated decision-making capabilities, enabling them to make informed decisions and propose appropriate courses of actions.
Its automation dimension mainly focuses on automating data flow and process execution through agent interactions, albeit limited to known data paths. 
However, the prototype overlooks crucial aspects such as financial and product flows, which are integral to a functional supply chain. 
For an effective ASC system, both intelligence and automation dimensions must be addressed, thereby enabling to handle a broader range of scenarios and challenges.

Meanwhile, as emphasised in \citet{xu2023autonomous}, the development of ASCs requires consideration of critical aspects extending beyond technological efforts.
These aspects include cyber security, data security and privacy, trustworthiness, and platform neutrality. 
Effectively addressing these aspects need collaborative efforts from all participants within the ASC ecosystem. 
While this initial proof-of-concept prototype may not fully addressed all of these aspects, its adoption of an MAS approach for integrating distributed entities represents a promising step toward addressing platform neutrality in the broader ASC domain.

\section{Conclusion and Future Work}\label{sec:conclusion}
This paper focuses on realising an autonomous supply chain (ASC) system, an recently emerging concept in the domain of SCM. 
Specifically, it introduces the agent-based autonomous supply chain (A2SC) through a multi-agent system (MAS) approach.

We have presented a methodology for A2SC analysis and design, which is adapted from a classic agent-oriented analysis and design framework, the Gaia methodology \citep{wooldridge2000gaia}. 
We have provided a detailed concrete case study, the autonomous meat supply chain, demonstrating the technical implementation of an A2SC system using the MAS approach.
Additionally, we presented a system architecture and a toolkit for developing A2SCs, offering guidelines for implementing similar decentralised systems.

This work marks an initial attempt to implement an ASC system, but there are certain limitations, as discussed in the previous section. 
Future efforts will mainly focusing on addressing these limitations, which include 
1) development of appropriate languages for agent communication and content representation for A2SC;
2) designing or introducing protocols capable of handling more complex scenarios, such as bargaining and multi-issue negotiations;
3) scaling up the problem, including expanding the number of stakeholders;
4) conducting experiments in less restricted setting; and 
5) equipping agents with intelligent decision making and planning capabilities.
Furthermore, we will explore the potential applications of emerging technologies and concepts within ASCs, such as Digital Twin (DT) \citep{sharma2022digital} and large language models (LLMs) \citep{bommasani2021opportunities, xu2023multi}.  
A2SCs can be viewed as a unique form of supply chain digital twins, which are not only virtual representations but also control centres for actual supply chains.
This perspective positions DT technologies and related concepts as promising avenues for advancing ASC development.
Recent progress in LLMs, such as ChatGPT \citep{openai2021chatgpt} and Llama \citep{touvron2023llama}, may enable the creation of AI agents that are truly capable of making decisions and planning on a more broad range of supply chain tasks with minimal human intervention.

In conclusion, this study is among the earliest explorations of implementing an ASC system with automated integrated processes by following a software engineering design methodology.
It demonstrates a promising pathway for effective ASC system development. 
The methodology for A2SC analysis and design, system architecture, toolkit for implementation, and agent roles and services model presented in this paper shed light on the future research and technical implementation of ASC systems in various scenarios.

\section*{Acknowledgement}
This work was supported by the Research England's Connecting Capability Fund (grant number: CCF18-7157): Promoting the Internet of Things via Collaboration between HEIs and Industry (Pitch-In) and the EPSRC Connected Everything Network Plus under grant EP/S036113/1.

\bibliography{references}

\pagebreak
\appendix

\section{Toolkit for Implementation}\label{appx:toolkit}
The tools and frameworks that were employed for developing the A2SC prototype are described below:
\begin{itemize}[nosep]

    \item {\it AEA Framework}:
    The AEA (Autonomous Economic Agent) framework \citep{minarsch2021autonomous} from Fetch.ai \footnote{\url{https://fetch.ai/}} technology stack is a Python-based toolkit for constructing autonomous agents known as AEAs. 
    These AEAs operate independently on behalf of their owners and autonomously execute actions to achieve predefined objectives.
    We used this framework to develop the agent network in the prototype.

    \item {\it OEF (Open Economic Framework)}:
    The OEF serves as a decentralised communication, search and discovery system for AEAs, providing protocols, languages, connection mechanisms that facilitate agents search, interaction discovery, and engagement within established social norms.
    It establishes a virtual {\it environment} where the inhabitants, AEAs, can operate and interact.
    In Fetch.ai technology stack, the AEA framework and the OEF represent the top two layers, positioned above the smart ledger layer.
    For this implementation, we exclusively utilised the AEA framework and the OEF to develop the prototype, as they are capable of functioning independently without the smart ledger.

    \item {\it REST and OpenAPI}:
    REST (REpresentational State Transfer) \citet{fielding2000architectural} is an architectural style for providing interface standards between decoupled components on the web.
    RESTful APIs are typically based on HTTP methods to access resources via URL-encoded parameters and the use of JSON or XML for data transmission. 
    All agents in the prototype were equipped with RESTful APIs, enabling them communicate with each other without the OEF and access external resources. 
    However, these agents need to expose their APIs to other in a standard manner. 
    The OpenAPI specification (OAS) \citep{openapi2021} was therefore adopted for describing these interfaces.
    The OAS defines a standard, programming language-agnostic interface description for RESTful APIs, allowing both humans and machines to discover and comprehend a web service without requiring access to source code and extra documentation. 
    In this prototype, we define interface description files (\texttt{.yaml}) by adhering to the OAS.

    \item {\it WebSockets}:
    WebSockets, which enable bi-directional, full-duplex communication, are well-suited for frequently updated or real-time applications. 
    We thus used WebSockets \citep{fette2011websocket} for services such as notification, monitoring, and agent chatting. 
    The above tools are relevant to backend development and architectural design.
    We will describe the tools/frameworks used for frontend development in this following paragraphs.

    \item {\it Django and Django Channels}:
    Django \footnote{\url{https://www.djangoproject.com/}} is a Python-based, fast, and easy-to-use web framework that follows the model-template-views (MTV) architectural pattern.
    It emphasises reusability and pluggability of components, which can serve as a hub to integrate heterogeneous components such as AEAs.
    Django, while versatile, is not well-suited for protocols require long-running connections, such as WebSockets, MQTT, and chatbots.
    Channels \footnote{\url{https://channels.readthedocs.io/en/stable/}} extends Django's capabilities to handle both HTTP and real-time applications.
    We used Django and Channels for developing web-based interfaces and integrating various UI elements into a browser-accessible portal.

    \item {\it Leaflet and OpenStreetMap}:
    Leaflet \footnote{\url{https://leafletjs.com/}} is an open-source JavaScript library designed for building interactive maps. 
    It offers lightweight, user-friendly, extendable, and cross-platform map creation capabilities.
    OpenStreetMap (OSM) \footnote{\url{https://www.openstreetmap.org}}, rather than other map services like Google Maps, was used as the map services due to its free of charge and sufficient effectiveness. 
    We utilised Leaflet in conjunction with OSM to provide map service for monitoring logistics activities in this prototype.

    \item {\it Chart.js}:
    Chart.js is a free, open-source, and simple-to-use JavaScript library for creating HTML-based charts. 
    We used Chart.js \footnote{\url{https://www.chartjs.org/}} to visualise real-time IoT data related to the ambient conditions of goods in this prototype.
\end{itemize}

In addition to these major tools, we incorporated complementary tools and frameworks such as Bootstrap \footnote{\url{https://getbootstrap.com/}}, JQuery \footnote{\url{https://jquery.com/}}, and Ajax.

\section{Implementation Details}\label{appx:implementation_details}
We adopted a loosely coupled architecture, in which the frontend interface and the backend agent network can be developed independently.
These two components were integrated into a unified system through the use of RESTful APIs and the WebSocket protocol. 
In total, we implemented {\it seven} distinct agents within the agent network, comprising one supplier, one wholesaler, two retailers, one logistics, two 3PL providers, and one admin.
These agents were individually developed using the AEA framework and collaborate by exchanging messages following predefined interaction protocols.

\begin{figure}
    \centering
    \includegraphics[width=\textwidth]{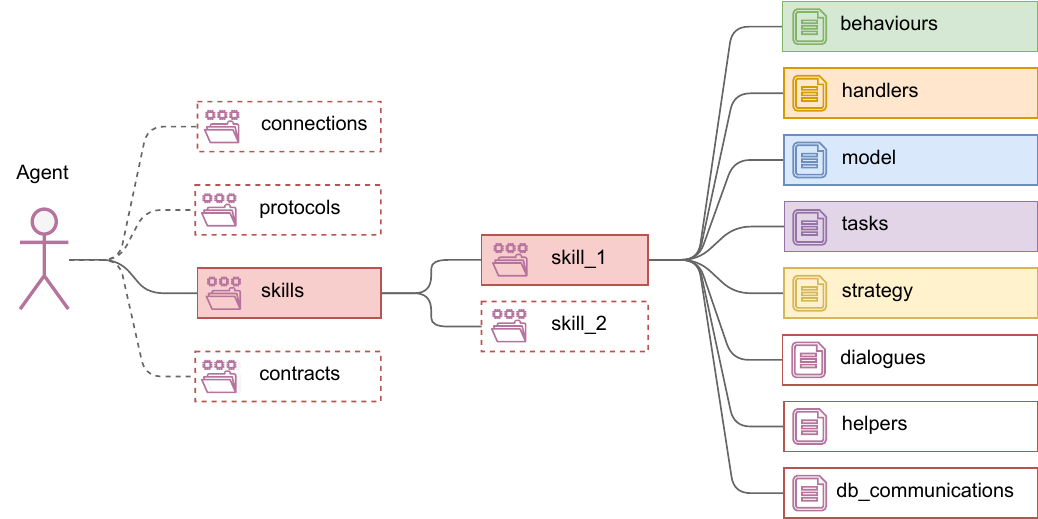}
    \caption{A visual representation depicting the directory structure of agents within this prototype.}
    \label{fig:agent_directory_structure}
\end{figure}

The AEA framework provides a base structure for agents, resulting in the directory structure employed for agent development in this prototype.
The agent directory structure is illustrated in \figureautorefname\ref{fig:agent_directory_structure}.
An agent directory is composed of four main packages for further development: \texttt{connections}, \texttt{contracts}, \texttt{protocols}, and \texttt{skills}. 
These packages are responsible for implementing agent-specific business logic. 
In out implementation, we used built-in implementations for \texttt{connections}, and \texttt{protocols}. 
Since this prototype does not include any features related to blockchain, we did not consider \texttt{contracts}, which typically contains wrappers for smart contract logic. 
Our primary focus was on \texttt{skills}, which define the specific services and capabilities of an agent.
An agent may possess multiple skills, each tailored to handle interactions with other agents under specific scenarios.
Skills are divided into types of actions: {\it reactive} and {\it proactive}, implemented through a set of handlers and behaviours, respectively.
Handlers are responsible for processing incoming messages following a specified protocol, constituting an agent's reactive behaviour.
Behaviours encompass self-initiated actions driven by internal logic rather than external messages.
Additionally, a skill incorporates data models for maintaining shared state within a skill, including description of the services that an agent with this skill can offer.

To manage CPU-bound or long-running executions, the framework provides the capability to create \texttt{tasks} within handlers and behaviours. 
We utilised this capability to handle long-executing tasks such as monitoring the delivery process.
Decision-making logic, such as assessing a proposal, is organised into a \texttt{strategy} file within a skill.
A skill may contain files related to agent dialogue management (\texttt{dialogues}), database connections (\texttt{db\_communications}), and miscellaneous utilities (\texttt{helpers}).
An agent's capabilities are configurable, allowing tasks such as configuring a skill or specifying a RESTful service.
Refer to the work by \citet{minarsch2021autonomous} for more details on the AEA framework.

The implemented agents can access two types of data sources: databases and sensors.
Agents such as supplier, wholesaler, and retailer manage their own inventory systems, which were simulated by updating their respective {\it inventory} table when products are inbound or outbound.
The 3PL agents require access to sensors to for delivery monitoring, which was simulated by periodically read the data that was pre-captured using the method described in \autoref{sec:data_preparation}.

The interface of this prototype was developed using Django and its plugin Channels.
WebSockets and RESTful APIs were used to exchange data and control between the web interface and the agent network, effectively connecting the frontend and backend of the system.
Real-time data, including vehicle location, agent messages, and system notifications, were transmitted to the interface using WebSockets, in which consumers was created to handle these data.
These data were then rendered using appropriate frontend techniques.
For instance, Chart.js was used for visualising IoT data, while OSM and leaflet were employed for tracking and monitoring the delivery vehicle. 
The agent chat room was developed from scratch without relying on specific libraries.
The ordering panel was implemented using Ajax, a set of client-side techniques that enable web applications to send and receive from a server (i.e., an agent) asynchronously without interfering with the display and behaviour of the existing page.
To enhance the layout and styling of the interface, Bootstrap and JQuery were utilised.
Further details about the tools used in developing the prototype are provided in \autoref{sec:toolkit}.

\section{Additional Figures}\label{appx:additional_figures}
\begin{landscape}
\begin{figure}[t]
     \centering
     \begin{subfigure}[t]{0.725\textwidth}
        \includegraphics[width=\textwidth]{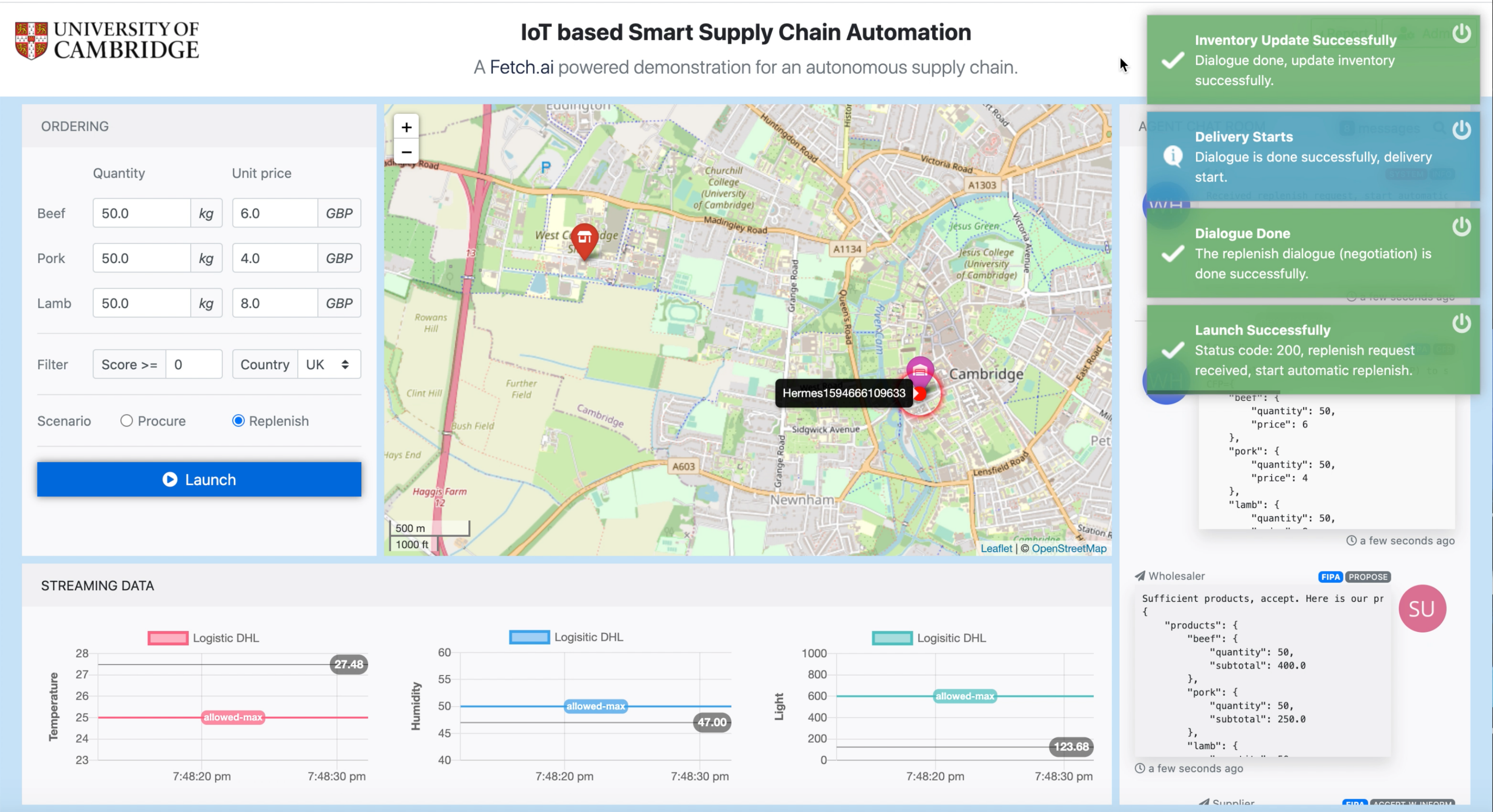}
        \caption{Replenishment: dialogue completed, delivery initiated. 
                 Relevant system notifications appear in the top right corner of the interface. 
                 On the map, you can see the locations of the supplier and wholesaler, which are the source and destination for the delivery with the tracking number: \texttt{Hermes1594666109633}.
                 The ongoing dialogues between the agents are displayed in the agent chat room on the right.}
        \label{fig:interface_replenish_start}
     \end{subfigure}
     \begin{subfigure}[t]{0.725\textwidth}
        \includegraphics[width=\textwidth]{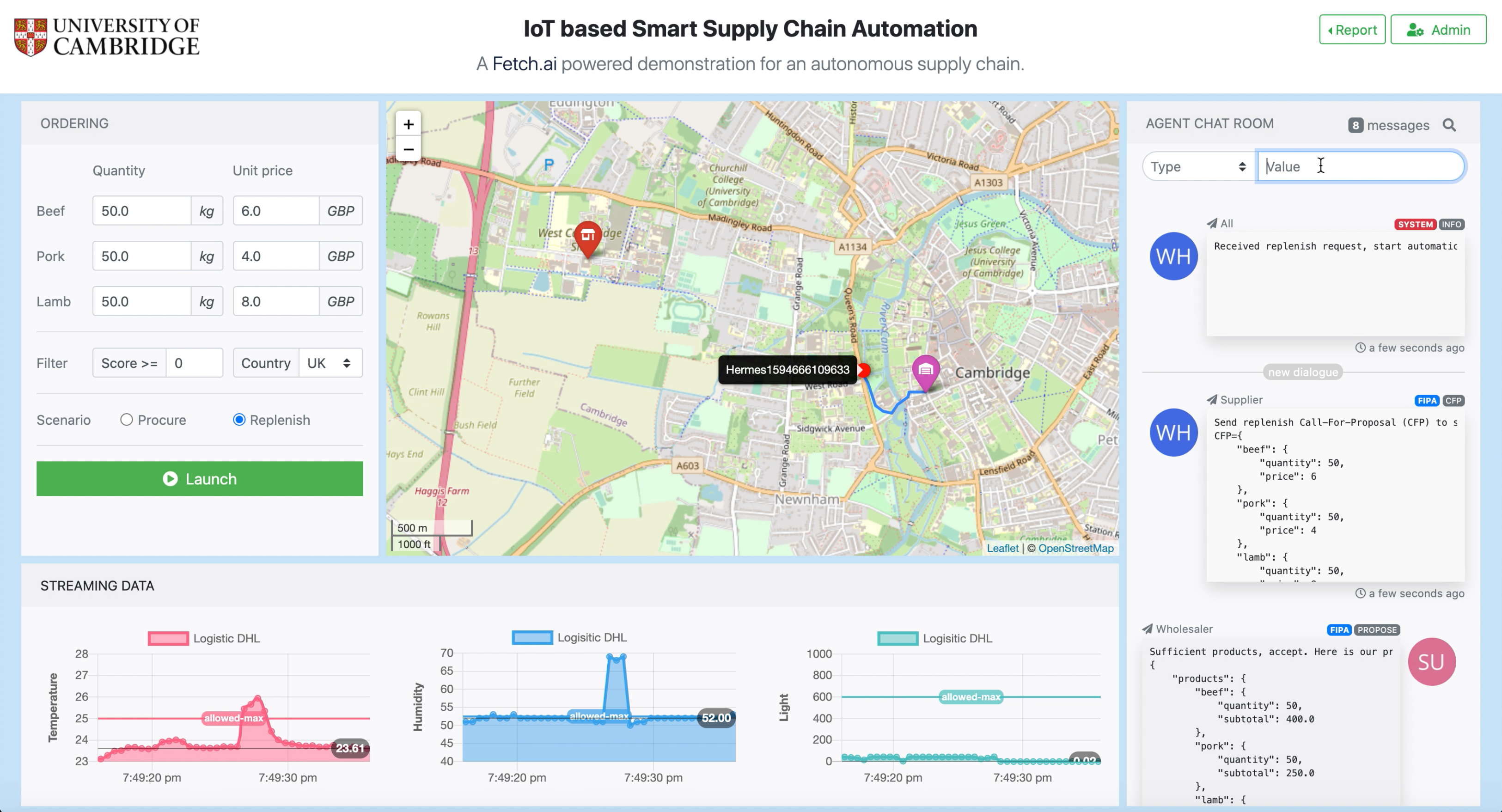}
        \caption{Replenishment: delivery in process.
                 The vehicle is current in transit, transporting the meat to its destination. 
                 Its real-time location is tracked and traced, represented by a blinking circle with a blue solid line on the map.
                 Additionally, the real-time ambient conditions of the meat, including temperature, humidity, and lightness, are visualised through the charts on the bottom area.}
        \label{fig:interface_replenish_delivery}
     \end{subfigure}
     \begin{subfigure}[t]{0.725\textwidth}
        \includegraphics[width=\textwidth]{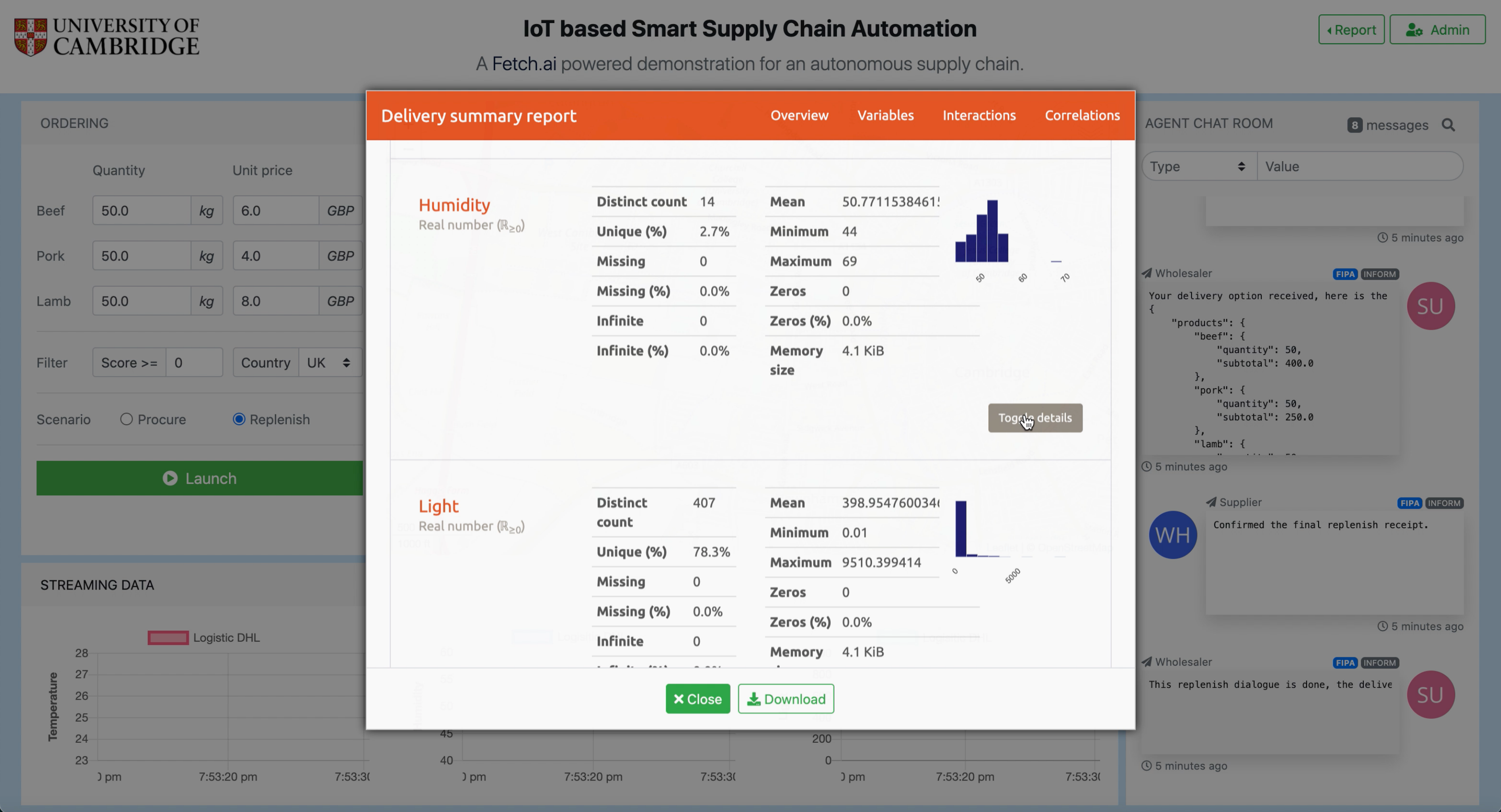}
        \caption{Replenishment: delivery completed, summary report available.
                 Once the products have been successfully delivered, a summary report generated from the collected the IoT data during the delivery journey is available to access. 
                 To view the report, simply click by the `Report' button located in the top right corner of the interface.
                 A portion of the summary report will appear in a scrollable pop-up window at the centre of the interface.}
        \label{fig:interface_replenish_summary}
     \end{subfigure}
     \begin{subfigure}[t]{0.725\textwidth}
        \includegraphics[width=\textwidth]{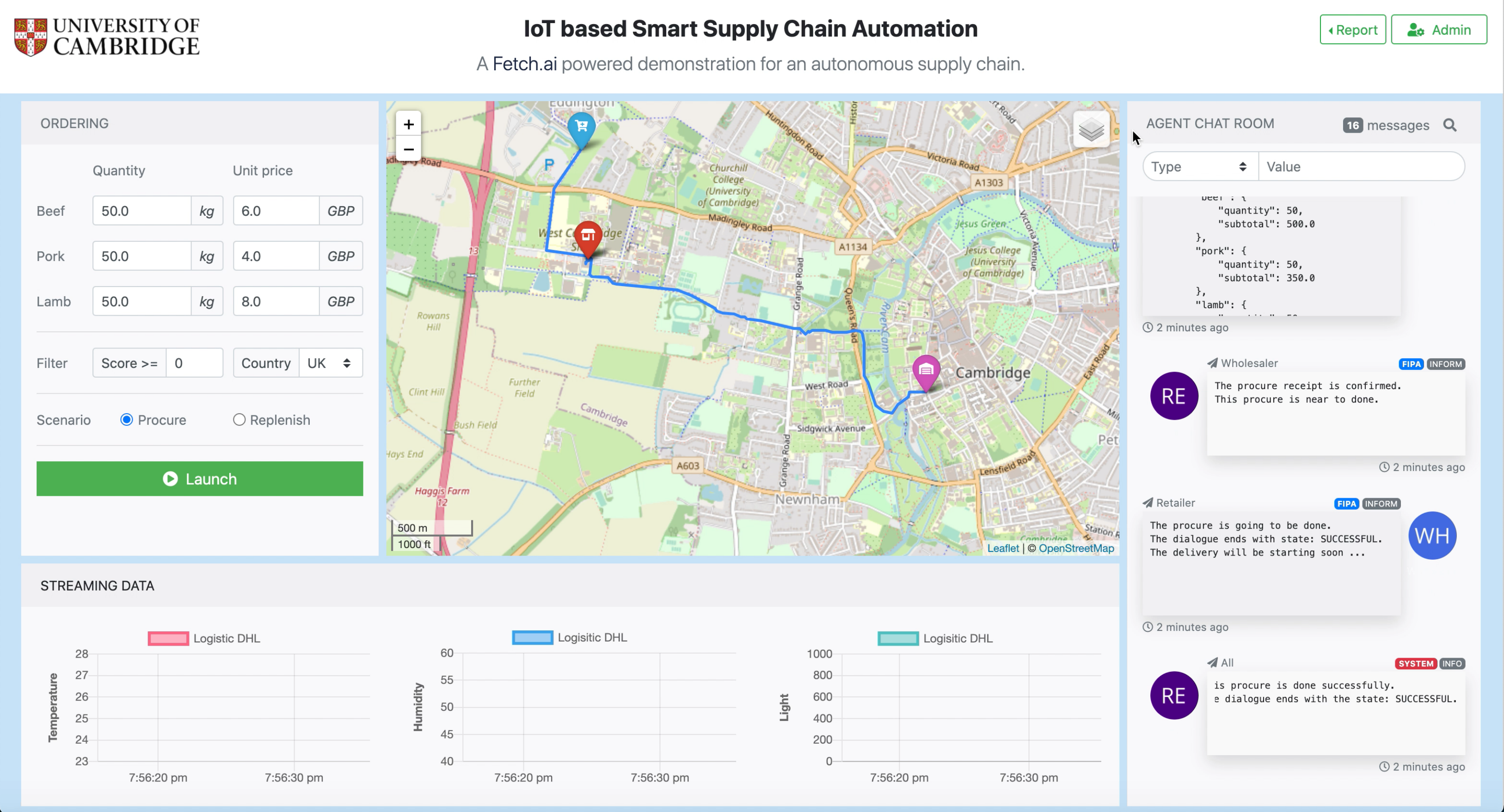}
        \caption{Replenishment and wholesale (procurement): delivery of both processes completed.
                 When the wholesale process is complete, the entire delivery route is displayed on the map, and a complete list of agent dialogue messages are shown in the agent chat room.
                 Their respective delivery summary reports can be accessed by either clicking the Report button or the light grey icon on the top right corner of the map.}
        \label{fig:interface_track_trace}
     \end{subfigure}
     \caption{Example screenshots of the prototype during the two processes: replenishment and wholesale (procurement).}
     \label{fig:interface_screenshots}
\end{figure}
\end{landscape}

\begin{landscape}
\begin{figure}[ht]
     \centering
     \begin{subfigure}[b]{0.36\textwidth}
         \centering
         \includegraphics[width=\textwidth]{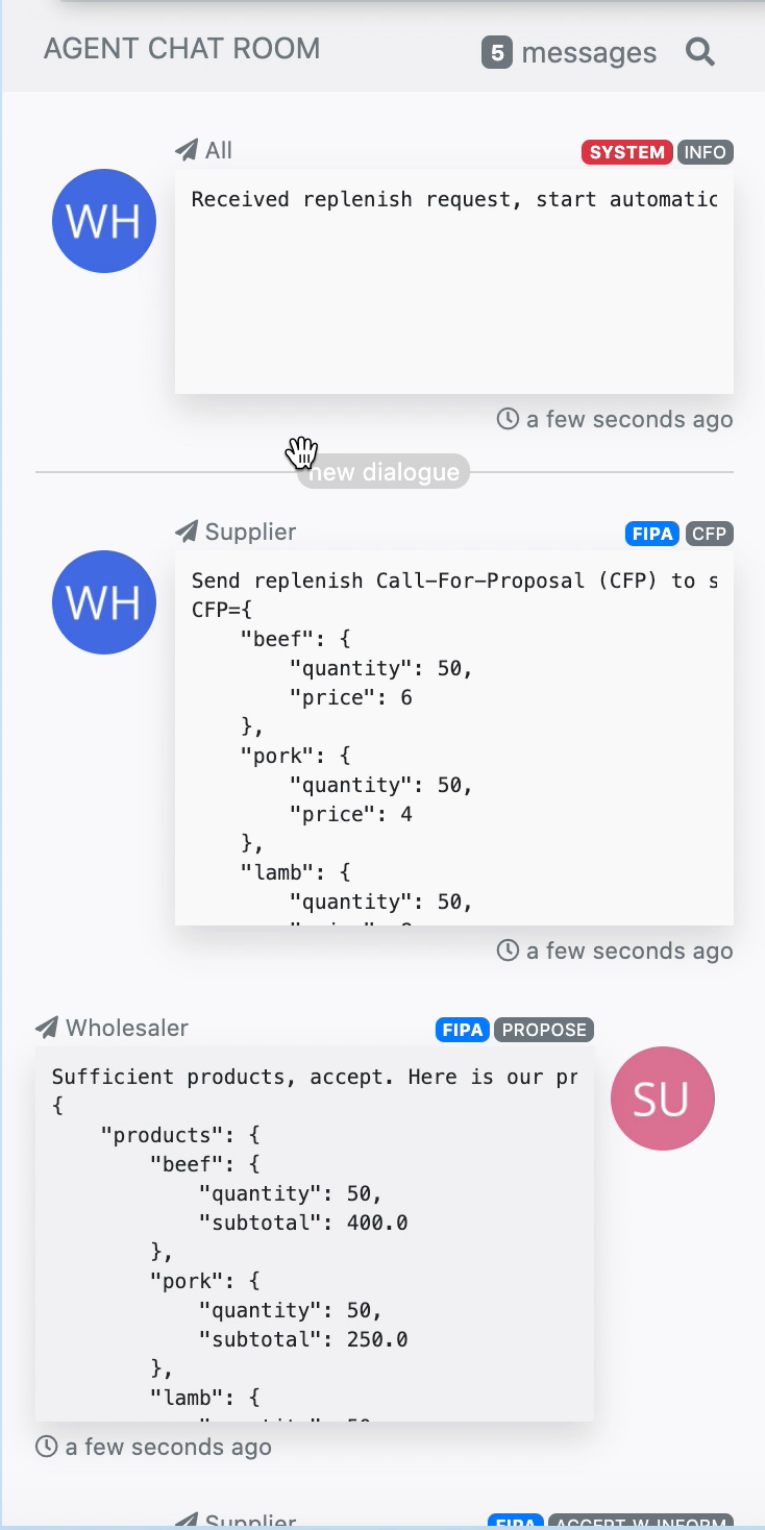}
         \caption{}
         \label{fig:excerpt_1}
     \end{subfigure}
     \begin{subfigure}[b]{0.36\textwidth}
         \centering
         \includegraphics[width=\textwidth]{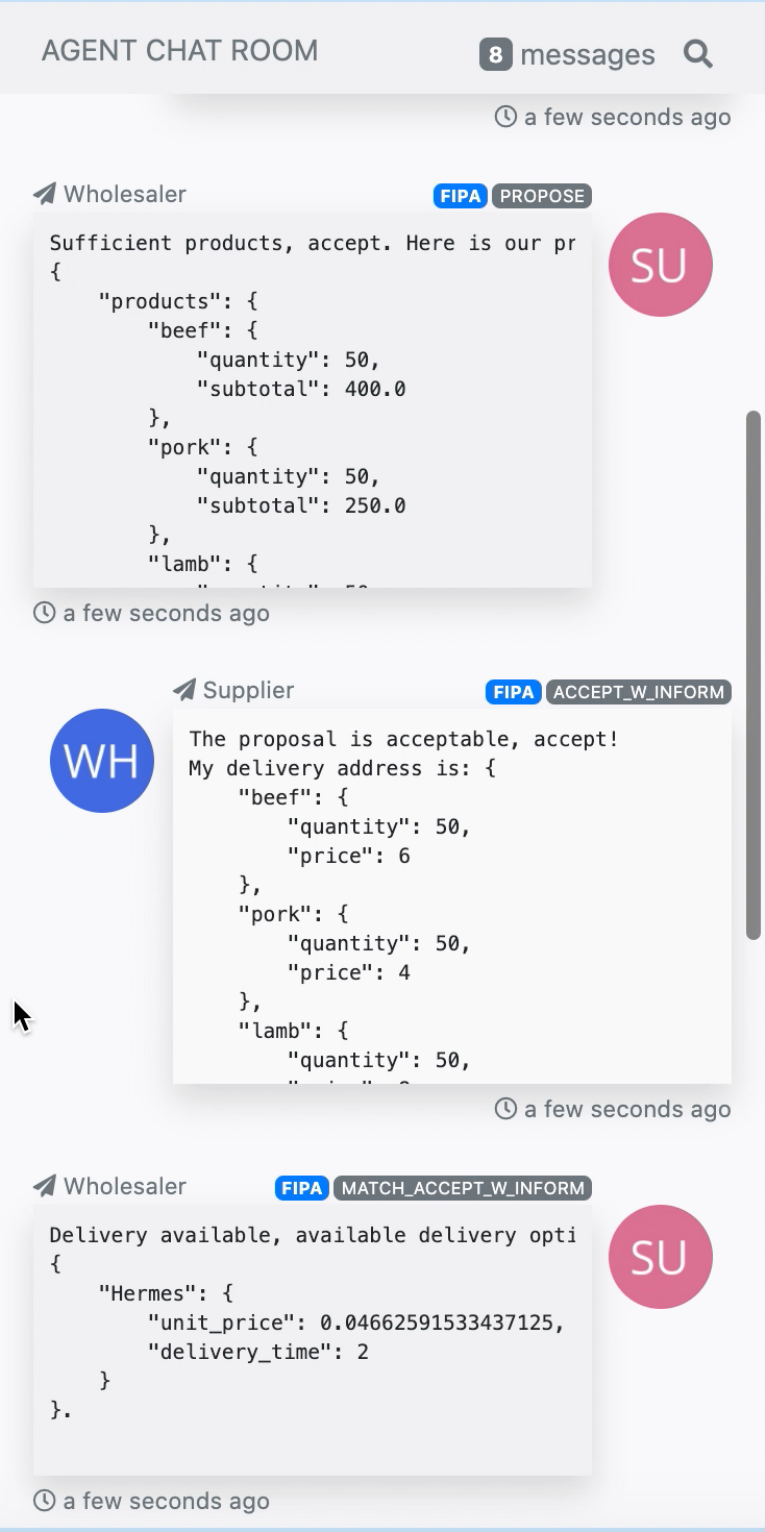}
         \caption{}
         \label{fig:excerpt_2}
     \end{subfigure}
     \begin{subfigure}[b]{0.3625\textwidth}
         \centering
         \includegraphics[width=\textwidth]{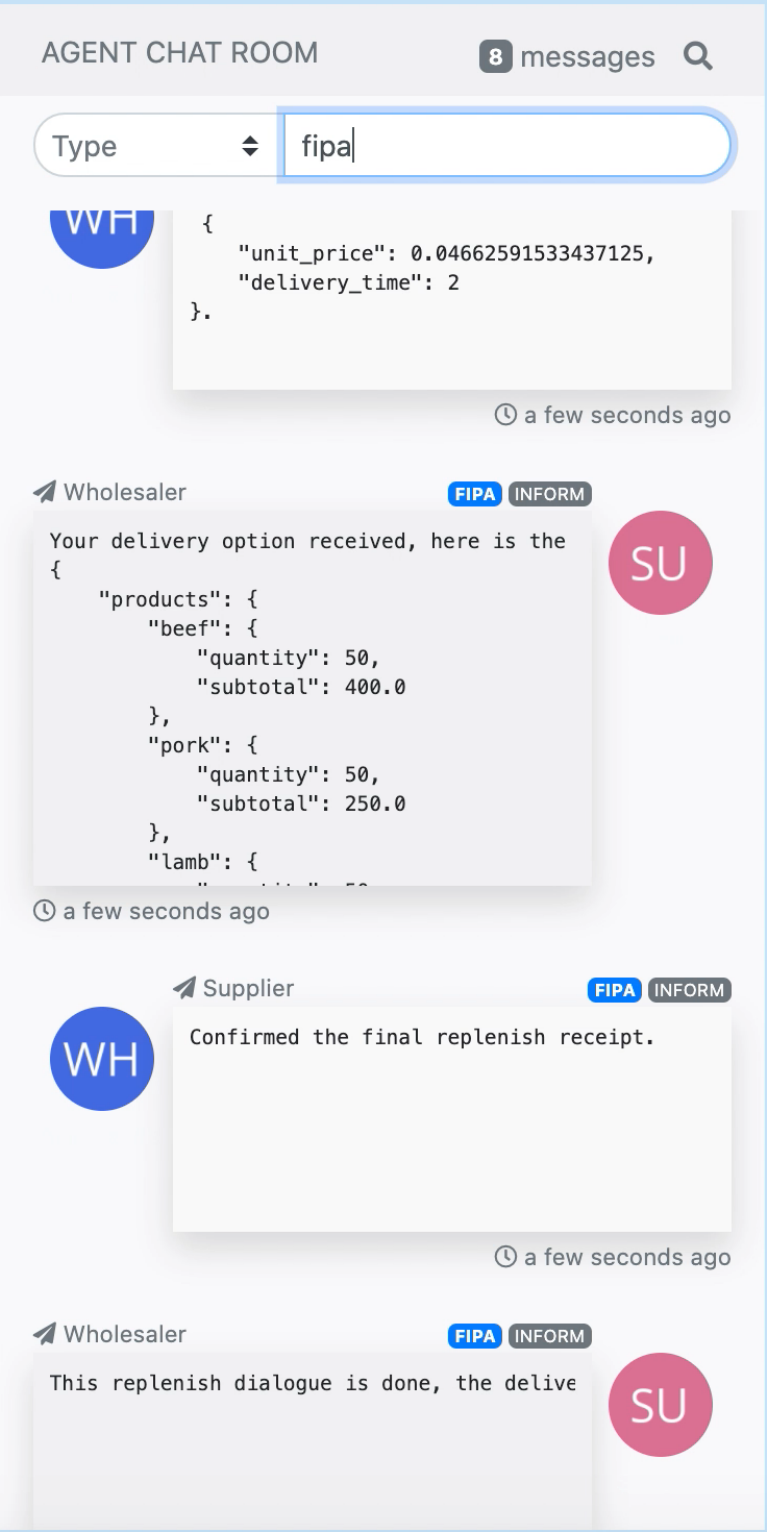}
         \caption{}
         \label{fig:excerpt_3}
     \end{subfigure}
     \begin{subfigure}[b]{0.3627\textwidth}
         \centering
         \includegraphics[width=\textwidth]{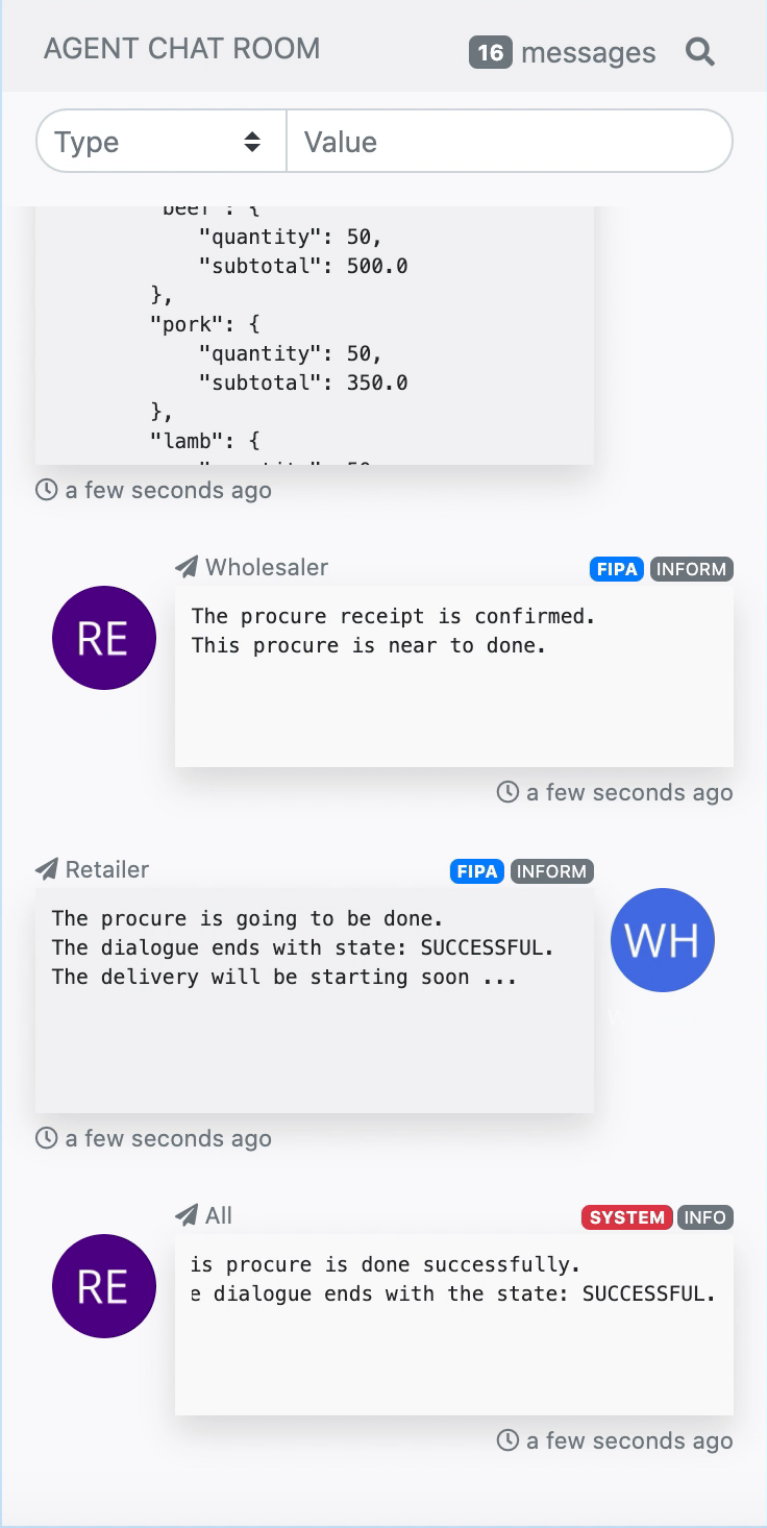}
         \caption{}
         \label{fig:excerpt_4}
     \end{subfigure}
     \caption{Agent chat message excerpts: (a), (b) and (c) are messages during the replenishment process; and (d) is for wholesale.}
     \label{fig:message_excerpts}
\end{figure}
\end{landscape}

\end{document}